\documentclass{aa}

\usepackage{graphicx}
\usepackage{txfonts}
\usepackage{siunitx}
\usepackage{xcolor}

\usepackage[normalem]{ulem}
\usepackage{amsmath}
\usepackage{hyperref}
\hypersetup{
  colorlinks=true,
  linkcolor=blue,
  urlcolor=blue,
  citecolor=blue
  }

\DeclareSIUnit\erg{erg}
\DeclareSIUnit\dyne{dyn}
\DeclareSIUnit\dn{DN}
\DeclareSIUnit\angstrom{\text {Å}}
\DeclareUnicodeCharacter{03B2}{$\beta$}

\DeclareSIUnit\gauss{G}

\begin{document}
\newcommand{\hrieuv}{HRIEUV\xspace}
\newcommand{\hrilya}{HRILy$\alpha$\xspace}
\newcommand{\fsieuv}{FSI 174 \xspace}

\newcommand{\seqa}{$\mathrm{S_1}$\xspace}
\newcommand{\seqb}{$\mathrm{S_2}$\xspace}
\newcommand{\seqc}{$\mathrm{S_3}$\xspace}

\newcommand{\eva}{$\mathrm{E_1}$\xspace}
\newcommand{\evb}{$\mathrm{E_2}$\xspace}
\newcommand{\evc}{$\mathrm{E_3}$\xspace}
\newcommand{\evd}{$\mathrm{E_4}$\xspace}
\newcommand{\eve}{$\mathrm{E_5}$\xspace}
\newcommand{\evf}{$\mathrm{E_6}$\xspace}
\newcommand{\evg}{$\mathrm{E_7}$\xspace}
\newcommand{\evh}{$\mathrm{E_8}$\xspace}

\newcommand{\evmethod}{$\mathrm{E_{9}}$\xspace}

\newcommand{\evlbeis}{$\mathrm{E_{10}}$\xspace}
\newcommand{\evl}{$\mathrm{E_{11}}$\xspace}
\newcommand{\evm}{$\mathrm{E_{11}}$\xspace}
\newcommand{\software}{\textit}

   \title{The nature of small-scale extreme ultraviolet solar brightenings investigated as impulsive heating of short loops in 1D hydrodynamics simulations}

   \author{A. Dolliou \inst{\ref{aff:ias},\ref{aff:mps}}
          \and J. A. Klimchuk \inst{\ref{aff:goffard}}
          \and S. Parenti \inst{\ref{aff:ias}}
        \and K. Bocchialini \inst{\ref{aff:ias}}
          }

    \institute{
        \label{aff:ias}{Université Paris--Saclay, CNRS,  Institut d'astrophysique spatiale, 91405, Orsay, France} \and
        \label{aff:mps}{Max Planck Institute for Solar System Research, Justus-von-Liebig-Weg 3, 37077 Göttingen, Germany} \and
        \label{aff:goffard}{Heliophysics Science Division, NASA Goddard Space Flight Center, Greenbelt, MD 20771, USA} 
    \\ \email{dolliou@mps.mpg.de}}

   \date{Accepted on 9 May 2025; published on}

 \abstract
{Small (400 to \SI{4000}{\kilo\meter}) and short-lived (10 to \SI{200}{\second}) extreme ultraviolet (EUV) brightenings, detected by the High Resolution Imager EUV (HRIEUV), have been found to be ubiquitous in the quiet Sun. Their contribution to coronal heating as well as their physical origin are currently being investigated.}
{We wish to determine whether models of short loops and impulsive heating are compatible with the results from observations. In particular, we used two models of loops with distinct thermal properties: cool ($T < $ \SI{1e5}{\kelvin}) and hot loops ($T > $ \SI{1e5}{\kelvin}).}
{We simulated the evolution of impulsively heated short loops, using the 1D hydrodynamics (HD) code HYDRAD. We computed the synthetic light curves of HRIEUV, four EUV channels of the Atmospheric Imaging Assembly (AIA), and five emission lines measured by the SPectral Imaging of the Coronal Environment (SPICE). We then compared the results from the synthetic light curves with observations. The aim was to reproduce the short delays observed between the intensity peaks of the light curves.}
{Cool loops subjected to impulsive heating are good candidates to explain the physical origin of the EUV brightenings. On the other hand, hot loops are not consistent with observations, except when they are subjected to especially strong impulsive heating.}

\keywords{Methods: numerical ; Sun: atmosphere ;
               }
    \titlerunning{The nature of small scale EUV solar brightenings investigated as impulsive heating of short loop}
    \authorrunning{Dolliou et al.}
   \maketitle

\section{Introduction}

The outer layer of the Sun's atmosphere, the corona, reaches temperatures above 1 MK. Magnetic reconnection and Magnetohydrodynamics (MHD) waves likely play a role in energy transfer and release from the photosphere to the corona. However, the exact processes that explain coronal heating are not yet clearly understood \citep[][]{Viall21,VanDoorsselaere_2020,Moriyasu_2004}. In active regions (ARs), observations in UV-EUV and X-ray showed signatures of impulsive heating \citep[\textit{e.g.}, ][]{Ugarte_2019,viall_survey_2017,Hinode_2019}, occurring especially at small scales \citep[\textit{e.g.}, ][]{Hudson1991,Crosby_1993,Hannah_2008,Aschwanden_2002}. As such, one of the main theories to explain coronal formation suggests that it is maintained by a large number of small-scale impulsive heating events called “nanoflares” \citep{Parker1988}. Despite being historically associated with magnetic reconnection, the term can now refer to any small-scale ($<$ \SI{e24}{\erg}) and impulsive energy releases, regardless of its physical origin \citep[][]{Klimchuk_2006,Klimchuk_2015}. The latter can be magnetic reconnection, MHD waves, or a combination of both, such as wave-induced reconnection \citep[][]{Sukarmadji_2024}. 

One approach to test the nanoflare theory is to detect the plasma response to nanoflares and to estimate its energy budget. As such, studies aimed to measure the impulsive emission associated with flares at the smallest possible energy scales, such as AR “microflares” ($\sim$\SI{e28}{\erg}) in X-ray \citep[][]{hannah_2019,Battaglia_2021}. Extreme ultraviolet imaging at high spatial and temporal resolution could reach even lower energy scales, down to $\sim$ \SI{e24}{\erg} in ARs \citep[]{Berghmans&Clette&Moses1998} and in the quiet Sun \citep[QS,][]{Berghmans_1998,Joulin_2016}. Hydrodynamic (HD) simulations of impulsively heated loops have also helped identify the properties of the heating (\textit{e.g.},  frequency, amplitude) and their observational signatures in the ARs \citep[][]{Barnes_2016a,Barnes_2016b,Cargill_2014,Viall_2015} and the QS \citep[][]{Upendran2021}.

The Solar Orbiter mission \citep[][]{Muller,Zouganelis2020}, launched in 2020, carries, among others, remote sensing instruments and has the particularity of regularly approaching the Sun up to 0.28 AU. As such, one of the mission objectives is to investigate the signatures of microflares at the smallest scales over a long period \citep[][]{Muller}. On May 30, 2020, the High Resolution Imager EUV (HRIEUV) of the Extreme Ultraviolet Imager \citep[EUI,][]{EUI_instrument}, on board Solar Orbiter, made one of its first high-cadence (\SI{5}{\second}) observations of the QS. During this \SI{4}{\min} sequence, \cite{Berghmans2021} reported the automatic detection of small (400 - \SI{4000}{\kilo\meter}) and short-lived (10 - \SI{200}{\second}) EUV brightenings. They are low-lying in the atmosphere \citep[1000 to \SI{5000}{\kilo\meter} above the photosphere,][]{Zhukov2021}, suggesting that they might originate from small magnetic structures. Similar events were also detected in later QS observations with HRIEUV, for instance in 2023 \citep[][]{Nelson_2024}. In fact, the regular Solar Orbiter observing plan (SOOP) (\textit{i.e.}, “R\_BOTH\_HRES\_HCAD\_Nanoflares”) showed that these events are present in large numbers and at all times in the QS. It implies that they are ubiquitous in the QS. 

The question remains as to whether these events are signatures of nanoflare heating. The first step in answering this question requires determining if they reach coronal temperatures ($T>$ \SI{1}{\mega\kelvin}), a necessary condition for them to contribute to coronal emission. 

Several approaches have been proposed to perform temperature diagnostics on the events: either relying on the statistics of a large number of events; or focusing on precise spectroscopic measurements on a few selected events. \cite{Dolliou_2023} chose the first approach, by analyzing the thermal behavior of the May 2020 event dataset using the EUV channels of the Atmospheric Imaging Assembly \citep[AIA,][]{Lemen2012} on board the Solar Dynamics Observatory \citep[SDO,][]{Pesnell2012}. Using the time lag analysis, they found that the events were characterized by short delays ($<$ \SI{12}{\second}) between the intensity peaks of pairs of AIA channels. Their main conclusion was that the event population was dominated by a TR component ($\sim$ \SI{e5}{\kelvin}), provided that the cooling time was long enough to be detected by AIA with \SI{12}{\second} of cadence. In parallel, \cite{Huang_2023} identify three events with the SPectral Imaging of the Coronal Environment \citep[SPICE,][]{Anderson_2020}, on board Solar Orbiter. These events show strong emission in the TR lines, while only a few are identified in the \ion{Ne}{VIII} ($\log{T} = 5.8$) and none in the \ion{Mg}{IX} ($\log{T} = 6.0$). These results are confirmed by \cite{Dolliou_2024}, who identify nine other QS events with SPICE and AIA. Furthermore, they report short delays between the intensity peaks of the HRIEUV, AIA, and SPICE light curves for most events. These delays are below \SI{6}{\second} for AIA and \SI{25}{\second} for SPICE. The authors also identify a single event with the EUV Imaging Spectrometer \citep[EIS,][]{Culhane_2007}, on board the Hinode spacecraft \citep[][]{Kosugi_2007}. Due to its wide coverage of coronal temperatures, working with EIS data provided evidence that most of the event's emission came from plasma at TR temperatures, suggesting that it does not significantly contribute to coronal heating. Lastly, \cite{Nelson_2023} identify Doppler velocities ($\approx$ \SI{23}{\kilo\meter\per\second}) with chromospheric and lower TR lines measured by the Interface Region Imaging Spectrograph \citep[IRIS,][]{DePontieu2014}. They conclude that some of the observed events share properties similar to those from Explosive Events \citep[EEs,][]{Teriaca_2004} already identified in the TR lines. In addition, the light curves of the chromospheric lines do not show typical behavior, suggesting different physical origins. This has also been thoroughly investigated, as we discuss below.

The physical origin of these events may be diverse, but reconnection is believed to play a role in at least some of them. Indeed, \cite{Kahil_2022} and \cite{Panesar_2021} find evidence of photospheric flux emergence and cancelation associated with some events. Furthermore, \cite{chen2021} performed simulations of the QS evolution using the 3D MHD code MURAM \citep{Vogler_2005,Rempel_2017}. The authors show that reconnection between magnetic flux tubes could induce impulsive EUV emissions similar to the events. \cite{Kuniyoshi_2024} propose an alternative mechanism to explain the origin of the events: impulsively generated Alfven waves, called “magnetic tornadoes." The latter have been observed for two decades in the chromosphere with ground-based telescopes \citep[][]{Bonet_2008,Wedemeyer-Bohm_2012}. While propagating upward along the field lines, the tornadoes can induce a plasma response in the upper atmosphere, which can be observed with EUV imaging \citep[][]{Zhang_2011}. Thus, very different physical processes can produce the impulsive emissions observed in HRIEUV.

The aim of our work is to use the radiation properties as a means to understand the formation of the events detected in HRIEUV. In particular, we wish to reproduce the short time delays observed between the HRIEUV, AIA, and SPICE intensity peaks \citep[][]{Dolliou_2023,Dolliou_2024} using a 1D HD model. For these, we assume that the emission comes from short loops with lengths (400 to \SI{4000}{\kilo\meter}) and heights (\SI{1000}{} to \SI{5000}{\kilo\meter}) similar to those derived from observations \citep[][]{Berghmans2021,Zhukov2021}. Short loops, with apex temperatures from \SI{e5}{\kelvin} to a few \SI{}{\mega\kelvin}, have already been observed in the TR and in the corona \citep[][]{Peter_2001,Winebarger_2013,Hansteen2014,Barczynski_2017,Feldman_1999}. Three-dimensional MHD simulations of the solar atmosphere evolution indicate that short loops can be locally heated to coronal temperatures through small-scale recurrent processes \citep{Skan_2023}. Because they are believed to be ubiquitous in the solar atmosphere, short loops, with their local and small-scale heating, can play a large role in maintaining the temperature of the low-altitude corona (below \SI{5000}{\kilo\meter}) above \SI{1}{\mega\kelvin}. This is especially likely in the QS, where only a few other large-scale structures can be found. While an impulsively heated short loop is a reasonable model for some events, we highlight that this assumption does not apply to events that appear on large-scale structures visible in HRIEUV \citep[\textit{e.g.}, Fig. B.1.b,][]{Dolliou_2024}. 

We use two different types of loop models to describe short loops, with distinct thermal properties: the so-called “hot” ($T_\mathrm{apex} > $ \SI{1e5}{\kelvin}) and “cool” loops ($T_\mathrm{apex} < $ \SI{1e5}{\kelvin}). Our goal is to understand how the choice of the loop model can impact its reaction to impulsive heating. Indeed, both models are dictated by different cooling terms: radiation and thermal conduction for hot loops, and radiation only for cool loops. We expect this difference to have an impact on the light curves' behavior. Thus, we want to understand which loop model can better reproduce the observational properties of the QS events detected by HRIEUV. In the following two paragraphs, we describe each model of loop in further detail.

Hot loops refer to the standard model used to describe a wide variety of loops above \SI{e5}{\kelvin}, from AR to the QS. \cite{Reale2014} proposed a review on the subject (see also Section \ref{sec:code:cl_hl}). Thermal conduction and radiative losses both play a role in the hot loop equilibrium state, but their relative magnitudes vary depending on the loop properties \citep[length, temperature, and density at the apex,][]{Cargill_1997}. For our models, the conductive cooling timescale ($\sim$ \SI{e2}{\second} to \SI{e3}{\second}) is lower or comparable to the radiative cooling timescale ($\sim$ \SI{e3}{\second}).

Cool loops were first theorized by \cite{Antiochos_1986}, in order to explain the observed increase of the the differential emission measure (DEM) at temperatures below \SI{e5}{\kelvin} \citep[for instance, see Fig.2 from][]{Parenti_2007}. For cool loops, thermal conduction is negligible, and radiation is in equilibrium with atmospheric heating throughout the loop (see Section \ref{sec:code:cl_hl}). Their equilibrium conditions depend on the choice of the radiative loss function, as discussed in \cite{Cally_1991} and in Appendix \ref{sec:annex:profile_cl}. In our case, cool loops can only exist in equilibrium with apex temperatures below \SI{1e5}{\kelvin}. \cite{Sasso_2012} proposed different types of cool loop models that exist for a wide range of radiative loss functions. In this work, we focus on the model originally described by \cite{Antiochos_1986}.

We now discuss our motivations for choosing loops in equilibrium in a cool and a hot state as starting models for our simulations. A basic property of EUV brightenings is that they suddenly become bright with short delays between different observing channels. The impulsive heating of an equilibrium cool loop is a natural explanation. The initial state is much denser than that of an equilibrium hot loop ($T\sim$ \SI{1}{\mega\kelvin}), and the emission measure difference is even greater because $\mathrm{EM} \propto n^2$. During a nanoflare, the temperature increases quickly with a timescale similar to that of the heating (e.g., \SI{10}{\second}). Density remains approximately unchanged during this short interval. Consequently, bright emission becomes visible in rapid succession in different observing channels as the plasma heats into their respective ranges of temperature sensitivity. The behavior of an impulsively heated hot loop is much different. It is initially faint because of the lower densities, even though the coronal temperature is within the range of sensitivity. The emission measure becomes substantial only after material evaporates into the loop. The brightest emission generally occurs when the heated plasma cools back down through the sensitivity range. This cooling is relatively slow and the delay between successive channels is correspondingly long.

Section \ref{sec:code} presents the code and the models used to simulate the evolution of impulsively heated short loops. Section \ref{sec:results} discusses the results obtained from the simulations, including the plasma parameters and the HRIEUV, AIA, and SPICE synthetic light curves (see Appendix \ref{sec:annex:forward_modelling}). In Section \ref{sec:discussion}, we compare the results with observations and discuss them in three main points. Finally, Section \ref{sec:conclusion} provides an overview and concludes this work.

\section{Hydrodynamic code and models}
\label{sec:code}
In Section \ref{sec:code:loop_geometry}, we introduce the semicircular geometry chosen for the loops. In Section \ref{sec:code:hydrad}, we present the 1D HD code used to simulate the evolution of the impulsively heated short loops. In Section \ref{sec:code:cl_hl}, we describe the properties of the cool and hot loops in equilibrium. Finally, in Section \ref{sec:code:groups_models}, we list the parameters chosen for each model of loop and impulsive heating.

\subsection{Loop geometry}
\label{sec:code:loop_geometry}
\begin{figure}
    \centering
    \includegraphics[width=\linewidth]{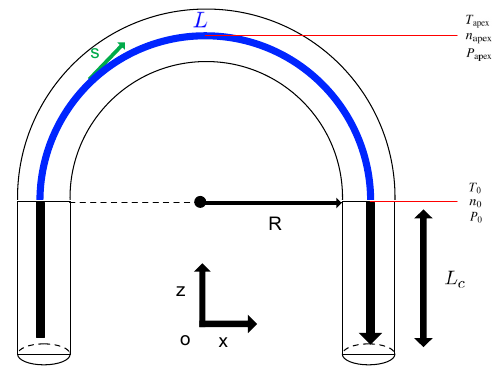}
    \caption{Geometry of a loop composed of a semicircular central part of length $L$ and apparent radius $R$ (in blue), with two vertical legs of length $L_\mathrm{c}$ (in black). Vector $\Vec{s}$ follows the magnetic axis of the loop, from left to right by convention. The electron temperature, density, and pressure are respectively indicated at the apex ($s = L_\mathrm{c} + L/2$) as $T_\mathrm{apex}$, $n_\mathrm{apex}$, and $P_\mathrm{apex}$ , and at the top of the vertical legs ($s = L_\mathrm{c}$ and $s = L_\mathrm{c} + L$) as $T_0$, $n_0$, and $P_0$. }
    \label{fig:code:semi_circular_geometry}
\end{figure}

Figure \ref{fig:code:semi_circular_geometry} is a diagram showing the geometry of the semicircular loops used in our models. They are made up of a central semicircular part of length $L$ (\SI{0.5}{} to \SI{5.5}{\mega\meter}) and two fixed length vertical legs $L_\mathrm{c} = $ \SI{10}{\mega\meter}. $R$ is the radius of the loop, and $s$ is the coordinate that follows the magnetic axis. $s$ increases from $0$ at the left footpoint to $2L_\mathrm{c} + L$ at the right footpoint. We also define an orthogonal coordinate system in the 2D plane ($x$, $z$), with $x$ the axis between the two footpoints, and $z$ the altitude.   

The chosen range of $L$ values is consistent with the apparent lengths of the events \cite[\SI{400}{} to \SI{4000}{\kilo\meter},][]{Berghmans2021}. On the other hand, $L_\mathrm{c}$ is set artificially larger than the measured height of the events above the photosphere \citep[1 to \SI{5}{\mega\meter},][]{Zhukov2021}. This is by design: large chromospheric legs provide a reservoir of plasma during the simulations in case of evaporation. Also, it allows for the perturbations originating from the loop's central part to be damped while they travel back and forth along the legs. However, we still aim for the amplitude of the gravitational acceleration vector $\vec{g}$ to be consistent with the estimated altitude of the events. To do so, $g$ is set to photospheric values for $z \leq$ \SI{9.5}{\mega\meter} and decreases with height above this limit.

Some of our models include loops with a semielliptical geometry, instead of a semicircular one (section \ref{annex:sec:groupIV_results}). In that case, the parameters described in Fig.\ref{sec:code:loop_geometry} still stand, with the exception of the apparent radius $R$. It is replaced by the semimajor axis $a$ and semiminor axis $b$ along the coordinates $x$ and $z$, respectively.

\subsection{1D hydrodynamic HYDRAD code with a two-fluid approach}
\label{sec:code:hydrad}

We use the 1D HD code HYDRAD with a two-fluid approach \citep{Bradshaw_2003,Bradshaw_Klimchuk_2011}, through its open-source version\footnote{\url{https://github.com/rice-solar-physics/HYDRAD}, consulted on 2024 August 5.}. This code computes the evolution with time of the plasma parameters, including the temperature $T$, density $n$, and pressure $P$ along the magnetic axis $s$ of the loop. The two-fluid approach implies that the evolution of the plasma parameters is computed separately for the electrons of mass $m_\mathrm{e}$ ($T_\mathrm{e}$, $n_\mathrm{e}$, $P_\mathrm{e}$) and the ion species of mass $m_\mathrm{i}$ ($T_\mathrm{i}$, $n_\mathrm{i}$, $P_\mathrm{i}$). The mass and momentum equations, along with the electron and ion energy conservation equations, are described in \cite{Bradshaw_Klimchuk_2011}. The ideal gas equations are used as closure.

The equation of electron energy conservation includes the atmospheric heating $H$ and the power loss due to radiations $Q_\mathrm{R}$ and electron conduction $C_\mathrm{e}$. The physical origin of $H$ is not known, but this term is necessary for the loop to reach an equilibrium. It is composed of a constant and uniform term $H_0$, representing the background heating, and an impulsive heating term with a Gaussian spatial deposition function:  

\begin{equation}
    \label{eq:code:H_ev}
    H_{\mathrm{ev}}(t, s) = A_{\mathrm{ev}}(t)\exp{-\frac{(s - s_\mathrm{ev})^2}{2\sigma_{\mathrm{ev}}^2}}.
\end{equation}

Here, $A_\mathrm{ev}(t)$ is the heating amplitude that increases and decreases linearly over time in a triangular shape, with $A_\mathrm{max}$ being defined as the maximum value. In our case, the increasing and decreasing periods of $A(t)$ are equal to \SI{5}{\second} each for all models. $\sigma_\mathrm{ev}$ is the standard deviation of the spatial deposition function, while $s_\mathrm{ev}$ is its center position. The power loss due to radiation is equal to the following:
\begin{equation}
\label{eq:code:q_r}
    Q_\mathrm{R} = n_\mathrm{e} n_\mathrm{i} \Lambda(T).
\end{equation}

We chose to represent the radiative loss function $\Lambda(T)$ (Eq. \ref{eq:code:q_r}) by a simplified piecewise continuous power law function \citep{Klimchuk_2008,Klimchuk_2010}, modified at temperatures below \SI{1e5}{\kelvin} to take into account the optical thickness of the \ion{H}{i} line \citep{McClymont_1983}:
\begin{equation}
\label{eq:code:lambda_t}
    \Lambda(T) = \left\lbrace \begin{matrix}
    \SI{8.87e-37}{} T^{3} & \mathrm{for}  & \log{T} \leq 5.0 \\
    \SI{8.87e-17}{} T^{-1}  & \mathrm{for} & 5.0 \leq  \log{T} \leq 5.67
    \\
    \SI{1.9e-22}{} & \mathrm{for}  & 5.67 \leq  \log{T} \leq 6.18
    \\
    \SI{3.53e-13}{} T^{-1.5} & \mathrm{for}  & 6.18 \leq  \log{T} \leq 6.55
    \\
    \SI{3.46e-25}{} T^{-0.33} & \mathrm{for}  & 6.55 \leq  \log{T} \leq 6.9
        \\
    \SI{5.49e-16}{} T^{-1.0} & \mathrm{for}  & 6.9 \leq  \log{T} \leq 7.63 \\
    \SI{1.96e-27}{} T^{0.5} & \mathrm{for}  & 7.63 \leq  \log{T}. 
    \end{matrix}\right.
\end{equation}

\subsection{Initial loops in equilibrium}
\label{sec:code:cl_hl}

\begin{figure*}
    \centering
        \includegraphics{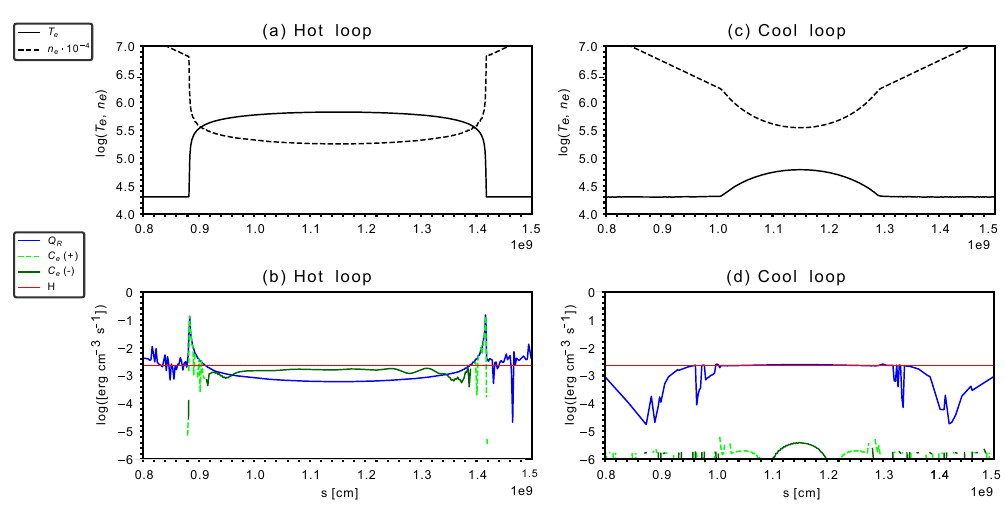}
    \caption{Examples of a cool (left column) and a hot (right column) loop in equilibrium. Top row: Electron temperature (\SI{}{\kelvin}) and density (\SI{e-4}{\per\centi\meter\tothe{3}}) profiles along the loop. Bottom row: Profiles of the electron energy conservation terms along the loop: $H=$ \SI{2.3e-3}{\erg\per\centi\meter\tothe{3}\per\second\tothe{1}} (red lines) is the uniform and constant heating term; $Q_\mathrm{R}$ is the power loss due to radiations; and $C_{\mathrm{e}} (+)$ (dotted light green lines) and $C_{\mathrm{e}} (-)$ (dark green lines) are the electron conduction, respectively, when they are a heating or a cooling term.}
    \label{fig:code:CL_et_HL_L3E8_Lc10E8_ntp}
\end{figure*}

In this section, we discuss the differences in thermal properties between the cool and hot loop models used in this work. We expect these differences to have an impact on how loops react to impulsive heating, implying different light curve behaviors. Figure \ref{fig:code:CL_et_HL_L3E8_Lc10E8_ntp} shows two initial cool and initial hot loops in equilibrium. It displays the profiles along $s$ of the electron temperature, density, and terms of the electron energy conservation. These terms include radiations $Q_\mathrm{R}$, electron conduction $C_\mathrm{e}$, the atmospheric heating $H$, here equal to $H_0$. Despite sharing the same geometry (semicircular) and length ($L=$ \SI{3}{\mega\meter}), the hot and cool loops show very distinct thermal properties.

For hot loops, the electron temperature and density profiles have a central coronal part surrounded by two transition regions (TR), with steep temperature and density gradients (Fig. \ref{fig:code:CL_et_HL_L3E8_Lc10E8_ntp}a). In the coronal part, radiation and electron conduction are cooling terms, while electron conduction becomes a heating term in the transition region (Fig. \ref{fig:code:CL_et_HL_L3E8_Lc10E8_ntp}b). In our simulations, hot loops in equilibrium are stable only for apex temperatures above \SI{1e5}{\kelvin} \citep[see also][]{Klimchuk_1987}.

For cool loops, the electron temperature and density profiles appear smooth and have no corona or TR (Fig. \ref{fig:code:CL_et_HL_L3E8_Lc10E8_ntp}c). The electron conduction ($\sim T^{5/2}\nabla T$) is negligible all along the loop because of the low temperatures and gradients of temperature. Therefore, coronal heating $H$ is compensated for by the power loss due to radiations $Q_\mathrm{R}$ only (Fig. \ref{fig:code:CL_et_HL_L3E8_Lc10E8_ntp}d).

The equilibrium conditions for cool loops set strong constraints on their maximal temperature and altitude above the photosphere. These limits are strongly tied to the shape of the radiative loss functions. Given our choice of radiative loss function following a power law $\Lambda(T) \sim T^b$, \cite{Cook_1989} set the condition $b > 2$ for cool loop equilibrium (see Appendix \ref{sec:annex:profile_cl} for more details). Therefore, in our simulations, the apex temperature of the cool loops has an upper limit equal to \SI{1e5}{\kelvin} (Eq. \ref{eq:code:lambda_t}). This limit was also given by \cite{Klimchuk_Mariska_1988} for the cool loops they described. They also provided an upper limit on the loop's altitude of \SI{2}{\mega\meter} above the chromosphere.

Lastly, we explain how we obtained a set of cool and hot loops in equilibrium. For the cool loops, we set the initial temperature, density, and pressure profiles described in \cite{Antiochos_1986}. The equilibrium was reached after a relaxation time of $\sim$ \SI{e4}{\second}. Hot loops in equilibrium were directly obtained from cool loops, by imposing a heating that brings the peak temperature above \SI{1e5}{\kelvin}. The cool loop then evolves into a hot loop (see section \ref{sec:results:transition_lc_hl} for more details). Again, a relaxation time of up to $\sim$ \SI{e4}{\second} is sometimes needed to reach equilibrium.

\subsection{Description of the model parameters}
\label{sec:code:groups_models}

In this section, we describe the models used and how the simulations were performed. Starting from a loop in an initial cool or hot state in equilibrium, we simulated its response to impulsive heating, from which we extracted the results presented in Section \ref{sec:results}. The initial simulation time was set to $t=$ \SI{0}{\second} by convention. At $t=$ \SI{100}{\second}, an impulsive heating was applied to the loop for a period of $\Delta t_\mathrm{ev} = $ \SI{10}{\second}. We chose to conduct an exploration of five specific parameters. The models are thus classified into five groups:

\begin{itemize}
    \item Group I (Table \ref{table:code:groupI}): The length $L$ of the loop central part varies from \SI{0.5}{} to \SI{5.5}{\mega\meter}. This group covers the range of apparent lengths measured for the events \citep[0.4 to \SI{4}{\mega\meter},][]{Berghmans2021}.
    \item Group II (Table \ref{table:code:groupII}): The pressure $P_0$ at the top of the vertical legs increases from 0.05 to \SI{0.5}{\dyne\per\centi\meter\tothe{2}}. The uniform atmospheric heating $H_0$ also varies as a consequence of the energy conservation on top of the vertical legs. This group covers different conditions of uniform heating and pressure in the solar atmosphere.
    \item Group III (Table \ref{table:code:groupIII}): The maximal amplitude of the impulsive heating $A_\mathrm{max}$ increases from 0.05 to \SI{0.5}{\erg\per\second\per\centi\meter\tothe{3}}. The aim of this group is to understand how the loop reacts to a wide range of impulsive heating, which might induce plasma flows.
    \item Group IV (Table \ref{table:code:groupIV}): The central parts of the loops have a semielliptical shape instead of a semicircular one. The semiaxis $a$ along $x$ increases from \SI{0.96}{} to \SI{2.5}{\mega\meter}. Loops in this group can be longer than those in group I, while being consistent with the observational constraints on their maximal height above the photosphere \citep[\SI{1}{} to \SI{5}{\mega\meter},][]{Zhukov2021}.
    \item Group V (Table \ref{table:code:groupV}): The deposition function of the impulsive heating is narrower ($\sigma_\mathrm{ev} = 0.1L$) than for other groups ($\sigma_\mathrm{ev} = 2L$). Furthermore, the center of the deposition function is either located at the loop apex or on its left side. The aim of this group is to determine if the location of the impulsive heating along the loop can have an impact on the induced plasma flows, and thus on the light curve behavior.
\end{itemize}

Each table provides the models with a name, along with a parameter set for the initial loop in equilibrium and for the impulsive heating. Each parameter set is associated with two models of loops in an initial cool state and in an initial hot state. There are two exceptions to this rule: no hot loop in equilibrium could be obtained for the shortest loops in group I, which are the ones from the models named m$_1$ and m$_2$ ($L=$ 0.5 and \SI{1}{\mega\meter}, see Table \ref{table:code:groupI}). In the figures and tables, the initial cool loop state is marked as "iCL", and the initial hot loop state as "iHL".

We wished to keep the impulsive heating similar between models as the loop size increased. To do so, we chose to keep constant the average volumetric energy deposited during the impulsive heating, defined as follows:

\begin{equation}
    E_{\mathrm{inj}} = \frac{1}{L}\int_{L_{\mathrm{c}}}^{L_\mathrm{c} + L} \int_{\Delta t_{\mathrm{ev}}} H_{\mathrm{ev}}(s, t) \ \mathrm{d}s \ \mathrm{d}t 
\end{equation}

For groups I to IV, we chose to subject the loop to an impulsive heating with a wide deposition function ($\sigma_{ev} = 2L$). The case of impulsive heating with a narrow deposition function localized at different positions along the loop is separately studied in group V. The reasons behind this choice are the following: the high temperature and pressure produced by a localized heating (\textit{i.e.}, nanoflare) expand very rapidly to fill the loop. By the end of the nanoflare, or shortly thereafter, the state of the loop is not greatly different from the “uniform” heating case, especially given the small loop lengths ($L/2 <$ \SI{2.25}{\mega\meter}). We checked this by looking into the evolution with time of the $T_{\mathrm{e}}$ and $n_{\mathrm{e}}$ profiles of m$_{19}$ (Fig. \ref{fig:annex:CL_HL_m19_ntv_v5}), which have a narrow deposition function for the impulsive heating ($\sigma= 0.1L$, see Table \ref{table:code:groupV}). In this case, the temperature starts increasing all along the loop \SI{10}{\second} after the start of the impulsive heating. Thermal conduction seems to be very efficient at spreading out the energy, along with a pressure-driven shock that develops at about t = \SI{105}{\second}. We estimate a shock speed of \SI{100}{\kilo\meter\per\second}.

Setting $\sigma = 2L$ for groups I to IV allows $E_{\mathrm{inj}} = $ \SI{0.99}{\erg\per\centi\meter\tothe{3}} to stay constant for groups I, II, and IV. For group III, $E_{\mathrm{inj}}$ increases with $A_\mathrm{max}$, which reaches up to \SI{0.5}{\erg\per\centi\meter\tothe{3}\per\second} for m$_{35}$. We can evaluate whether such impulsive heating is likely to happen in the QS. For m$_{35}$, the injected energy $E_{\mathrm{inj}}$ is equal to \SI{2.4}{\erg\per\centi\meter\tothe{3}}. If we equate this with the free magnetic energy in a stressed field, the stress component is \SI{7.9}{\gauss}. Assuming a shear angle of 20°, the stress component can be no more than 35\% of the total magnetic energy \citep[][]{Klimchuk_2015}. Thus, the total magnetic field must be around \SI{24}{\gauss}, which is high, but not unreasonable in the QS \citep[\textit{e.g.}, $B\approx$ \SI{12}{\gauss} at TR heights,][]{Rodriguez_Gomez_2024}.

The final goal was to compare the simulation results with those from observations, in order to constrain the parameters of the models. In that regard, we synthesized the light curves of HRIEUV, four EUV channels of AIA (\SI{171}{\angstrom}, \SI{131}{\angstrom}, \SI{193}{\angstrom}, \SI{211}{\angstrom}), and five emission lines measured by SPICE (\ion{C}{iii} \SI{977.03}{\angstrom}, \ion{O}{iv} \SI{787.72}{\angstrom}, \ion{O}{vi} \SI{1031.93}{\angstrom}, \ion{Ne}{viii} \SI{770.42}{\angstrom}, \ion{Mg}{ix} \SI{706.02}{\angstrom}). The response and contribution functions of all channels and lines are displayed in Fig. \ref{fig:annex:gofnt}. While most EUV channels of AIA have a sensitivity peak at coronal temperatures, most lines measured by SPICE are emitted by plasma at TR temperature. Only the \ion{Mg}{ix} \SI{706.02}{\angstrom} line is emitted by plasma at \SI{1}{\mega\kelvin}, but this line is weak in the QS. The intensities are integrated over equidistant vertical line of sights (LOS) parallel to the $z$ axis (Fig. \ref{fig:annex:m4_Tene}). In observations, the light curves are generally obtained by averaging the intensities over a spatial region around the event \citep[][]{Dolliou_2024,Huang_2023}. Thus, we assumed the short loop's emission not to be resolved. This is why we computed the light curves by averaging the intensities over all of the LOS, at each time step. More details on the forward modeling method are given in Appendix \ref{sec:annex:forward_modelling}. The light curves computed for the groups I and III models are shown in Section \ref{sec:results:lc}. For the other groups, the light curves are shown in Appendices \ref{annex:sec:groupII_results} to \ref{annex:sec:groupV_results}.

\section{Results}
\label{sec:results}

We now discuss the simulation results. In Section \ref{sec:results:transition_lc_hl}, we discuss the conditions for a cool loop to evolve into hot loops. Section \ref{sec:results:temperature_loop} shows the electron temperatures and densities reached by the apex of the loops before and after impulsive heating. Section \ref{sec:results:lc} displays the synthesized light curves. Finally, section \ref{sec:results:flows} uses two models of interest (m$_4$, and m$_{35}$) to describe the impact to the light curves of high density effect and plasma flows induced by the impulsive heating. In the following, we focus on the results of the groups I and III models. The results of the groups II, IV, and V models were redundant and are given in Appendix \ref{annex:sec:results_groups}.

\subsection{Conditions for the evolution of a loop from a cool into a hot state}
\label{sec:results:transition_lc_hl}

\begin{figure*}
    \centering
    \includegraphics{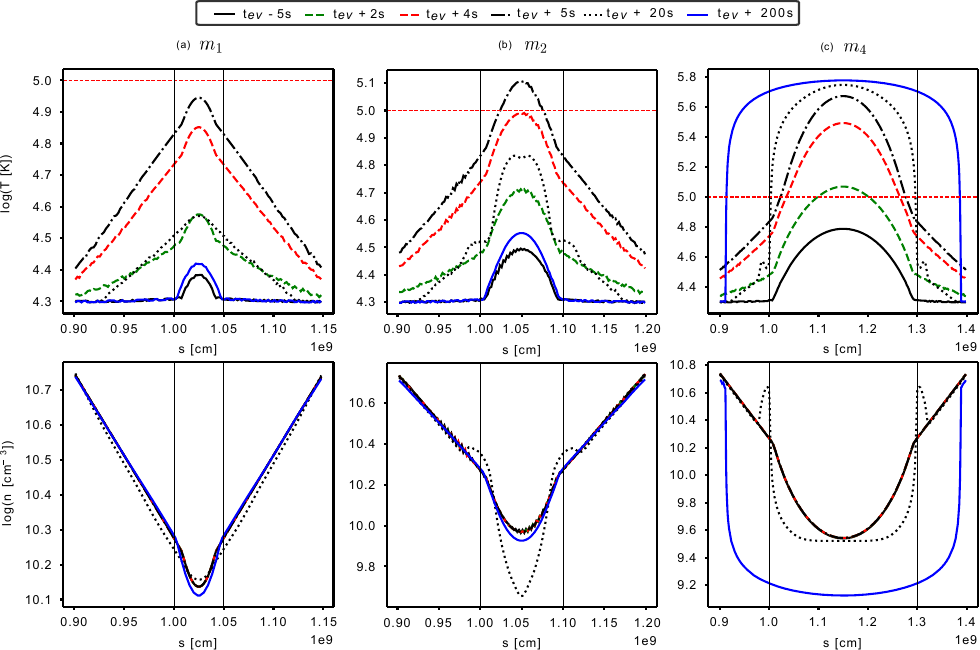}
    \caption{Evolution of the electron temperature (top row) and density (bottom row) profiles with time for the simulations with the models (a) m$_1$, (b) m$_2$, and (c) m$_3$ ($L=$ \SI{0.5}{}, \SI{1}{} and \SI{3}{\mega\meter}). The starting time of the impulsive heating is set at $t_{\mathrm{ev}} = $ \SI{100}{\second}. The vertical black lines indicate the basis of the loop's central part and the horizontal dotted red line shows the values $\log{T} = 5.0$. $t_{\mathrm{ev}} +$ \SI{2}{\second} (green) and $t_{\mathrm{ev}} +$ \SI{4}{\second} (red), which are the closest recorded times when m$_4$ and m$_2$ reach apex temperatures above $\log{T} = 5.0$.}
    \label{fig:results:m124_these_nt_evolution_v3}
\end{figure*}

In this section, we show how a loop in an initial cool state can evolve into a hot state, when the apex temperature is increased above $\log{T} = 5.0$. The latter is the upper limit of the apex temperature for a cool loop in equilibrium (Section \ref{sec:code:cl_hl}). As an example, we present the results from three group I models, namely m$_1$, m$_2$ and m$_4$, with an increasing length from $L=$ \SI{0.5}{} to \SI{3}{\mega\meter} (Table \ref{table:code:groupI}).

Figure \ref{fig:results:m124_these_nt_evolution_v3} shows the evolution with time of the electron temperature and density when cool loops of different lengths are subjected to the same impulsive heating. The equilibrium state of cool loops with increasing length reach higher apex temperatures (from $\log{T} =$ 4.4 to 4.8) and lower apex densities (from $\log{n} = $ 10.1 to 9.6). Therefore, the maximal temperature reached after a similar impulsive heating increases with the loop length. We can distinguish three cases where the loop does not reach (m$_1$), barely reaches (m$_2$) and reaches well above (m$_4$) the upper limit temperature for cool loops in equilibrium ($\log{T} = 5.0$).

The apex temperature of m$_1$ (Fig. \ref{fig:results:m124_these_nt_evolution_v3}a) stays below $\log{T} = 5.0$ and the cool loop does not evolve into a hot loop. Instead, the temperature and density profiles return to their original states after a time of about \SI{400}{\second}. On the other hand, the apex temperature of m$_4$ reaches values well above $\log{T} = 5.0$ (Fig. \ref{fig:results:m124_these_nt_evolution_v3}c). Above $\log{T} = 5.0$, the radiative loss function decreases with temperature (Eq. \ref{eq:code:lambda_t}), which accelerates the increase of the temperature even more. The cool loop then evolves into a hot loop. In our case, the process is irreversible as long as $H_0$ was kept constant. Model m$_2$ is a special case, as the loop apex temperature barely reaches values above $\log{T} = 5.0$ (Fig. \ref{fig:results:m124_these_nt_evolution_v3}b). In that case, the high density still provides important cooling through radiations (Eq. \ref{eq:code:q_r}). The latter is then sufficient to decrease the temperature of the loop below $\log{T} = 5.0$. For both models, m$_1$ and m$_2$, the temperature and density profiles eventually return to their original state. Thus, the loops return to their initial cool loop state in equilibrium.

Also, no hot loop state in equilibrium could be formed for models m$_1$ and m$_2$, regardless of how strong or localized the impulsive heating is. Indeed, these loops appear to be too short to reach a stable hot loop equilibrium. To test this hypothesis, we applied a stronger impulsive heating ($A_{\mathrm{max}}$ = \SI{0.5}{} and \SI{1.0}{\erg\per\centi\meter\tothe{3}\per\second}) to the initial cool loop of m$_2$, with a wide and a narrow deposition function ($\sigma_{\mathrm{ev}} = 2L$ and $0.1L$). For $\sigma_{\mathrm{ev}} = 0.1L$, we separate the cases where the impulsive heating is localized in the center of the loop, or near the left footpoint. Figure \ref{fig:annex:v5_referee_nt_evolution} displays the evolution with time of the temperature and density profiles, for all three cases. Despite reaching temperatures above $\log{T} = 5.0$, the loop eventually returned to its initial cool state for all cases, and could not reach a hot state in equilibrium. This is consistent with \cite{Klimchuk_1987}, who found that hot loops were unstable below a given height, equal to $z=$\SI{1}{\mega\meter} in their model.

One might ask whether a short loop with localized background heating – rather than our uniform background heating – might allow a stable hot equilibrium for the end state. We doubt this would be the case if the localized heating were at the apex. This is because the reasons for instability with uniform heating also apply with apex heating. A cool equilibrium likewise will not be possible if the localized heating is sufficiently strong that the final temperature would exceed \SI{0.1}{\mega\kelvin}. Thus, the loop will undergo a limit cycle of heating and cooling, never reaching a stable equilibrium. If the localized background heating is instead at low altitudes in both legs (symmetric heating), the loop will be in a state of thermal nonequilibrium, again undergoing an endless cycle of heating and cooling. The only possibility of a stable hot equilibrium is if the heating is asymmetric. In that case steady flows from one footpoint to the other – sometimes referred to as siphon flows – are an important part of the energy balance \citep[][]{Klimchuk_Luna_2019}.

\subsection{Temperature and density reached after impulsive heating}
\label{sec:results:temperature_loop}    

\begin{figure*}
    \includegraphics{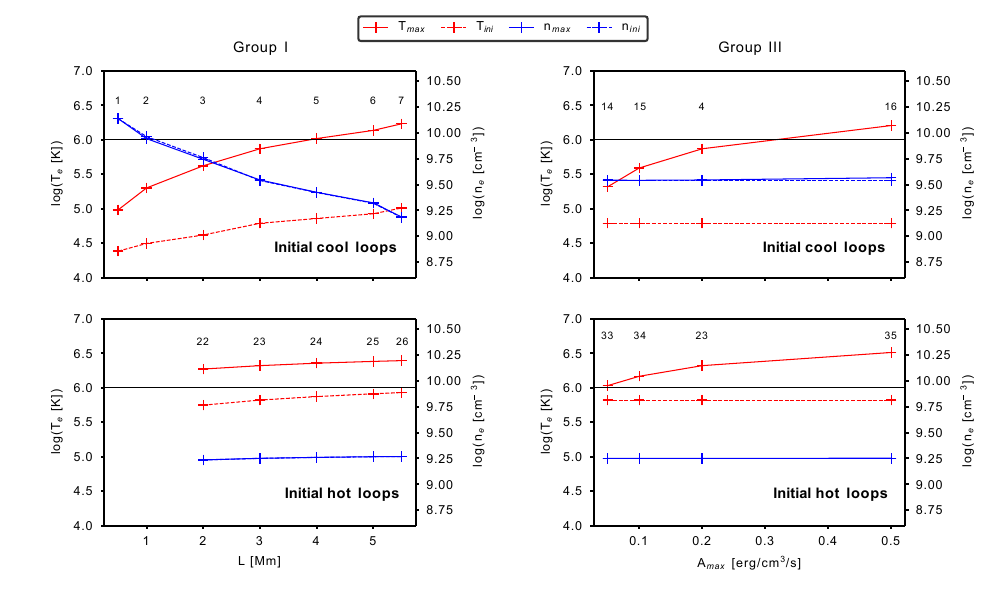}
    \caption{Electron temperature $T_\mathrm{max}$ and density $n_\mathrm{max}$ at the loop center ($s = L_\mathrm{c} + L/2$) at the time of the temperature peak after the impulsive heating. Prior to the impulsive heating, the initial values $T_\mathrm{ini}$ and $n_\mathrm{ini}$ at the loop center are provided for the loops in equilibrium. Results for groups I (left column) and III (right column) are shown for initial cool (upper row) and hot loops (lower row). The X-axis is the changing parameter associated with each group ($L$ or $A_\mathrm{max}$). Model numbers given in tables \ref{table:code:groupI} and \ref{table:code:groupIII} (\textit{i.e.}, m$_1$ to m$_{35}$) are displayed on top of their respective data points (crosses). The horizontal black line delimits the \SI{1}{\mega\kelvin} temperature, above which the heating is likely to contribute to coronal heating.}
    \label{fig:results:fig1_tmin_tmax}
\end{figure*}

\begin{figure*}
    \centering
    \hspace*{-1.3cm}
    \includegraphics{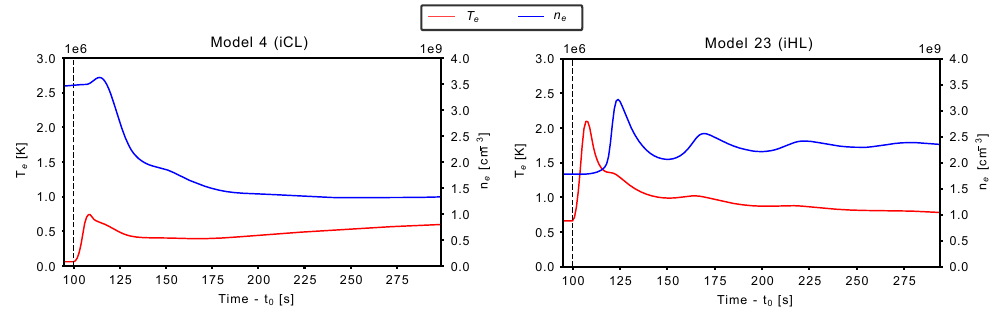}
    \caption{Evolution with time of the electron temperature and density at the loop center ($s=L_{\mathrm{c}} + L/2$) for two models with an initial cool loop (m$_4$) and an initial hot loops (m$_{23}$) of similar length ($L = $ \SI{3}{\mega\meter}). The dotted vertical line indicates the starting time of the impulsive heating.}
    \label{fig:results:m4_m23_n_t_apex}
\end{figure*}

Figure \ref{fig:results:fig1_tmin_tmax} shows the electron temperature and density in the center of the loop ($s =L_\mathrm{c} + L/2$). The values are given before the impulsive heating ($T_\mathrm{ini}, n_\mathrm{ini}$) and at the time of the temperature peak ($T_\mathrm{max}, n_\mathrm{max}$). We focus here on the results of the groups I and III models. Those from the other groups are shown in Fig. \ref{fig:annex:fig1_tmin_tmax}. In Section \ref{sec:discussion}, $T_\mathrm{max}$ and $n_\mathrm{max}$ are used to estimate the emission of the \ion{Mg}{ix} \SI{706.02}{\angstrom} line, and to compare it with the upper limits set by SPICE observations. Furthermore, measuring $T_\mathrm{max}$ can help determine if the loop is likely to contribute to coronal emission, as it requires that $T_\mathrm{max}>$ \SI{1}{\mega\kelvin}.

With regard to impulsively heated cool loops, only a fraction of them reaches coronal temperatures, defined as $T >$ \SI{1}{\mega\kelvin}. This includes those with the larger lengths ($L \geq$ \SI{4}{\mega\meter}) or those that are subjected to a sufficiently strong impulsive heating ($A_\mathrm{max} = $ \SI{0.5}{\erg\per\centi\meter\tothe{3}\per\second}). On the other hand, all models with impulsively heated initial hot loops reach coronal temperatures. This is because loops in an initial hot state are hotter and less dense than those in an initial cool state. For instance, given a similar length ($L=$ \SI{3}{\mega\meter}), the loop in an initial cool state of m$_4$ is cooler and denser ($\log{T_\mathrm{ini}} = 4.7$, $\log{n_\mathrm{ini}} = 9.5$) than those of m$_{23}$ in an initial hot state ($\log{T_\mathrm{ini}} = 5.7$, $\log{n_\mathrm{ini}} = 9.25$). When subjected to similar impulsive heating, the loop in m$_\mathrm{4}$ will reach a lower temperature ($\log{T_\mathrm{max}} = 5.9$) compared to the loop in m$_\mathrm{23}$ ($\log{T_\mathrm{max}} = 6.4$).

For all models, $n_\mathrm{max} \approx n_\mathrm{ini}$. This is because the increase in temperature is faster ($<$\SI{20}{\second} after the start of the impulsive heating) than the decrease in density for cool loops (from $\sim$ \SI{e1}{} up to \SI{e2}{\second}). The latter is due to the evolution of the loop from a cool state into a hot state, after impulsive heating (Section \ref{sec:results:transition_lc_hl}). For hot loops, density increases from evaporation, and this is slow compared to nanoflare heating and thermal conduction-dominated cooling that follows. The density maximum occurs after the temperature maximum. Also, $T_\mathrm{ini}$ and $n_\mathrm{ini}$ both increase with length for hot loops (group I), in accordance with equilibrium scaling laws \citep[$T_\mathrm{max} \sim 1.4\cdot10^3(P L)^{1/3}$,][]{Rosner_1978}. For cool loops in equilibrium, on the other hand, $T_\mathrm{ini}$ increases with $L$ while $n_\mathrm{ini}$ decreases with $L$.

Figure \ref{fig:results:m4_m23_n_t_apex} shows in more detail the evolution with time of the electron temperature and density at the loop center, for two models with loops in an initial cool (m$_4$) and hot state (m$_{23}$) of similar length ($L = $ \SI{3}{\mega\meter}). In m$_4$, the peak in temperature is reached in about \SI{10}{\second}, followed by a slower decrease of the density (over more than \SI{50}{\second}). The radiative losses decrease with $T_{\mathrm{e}}$ for $T_{\mathrm{e}} >$ \SI{0.1}{\mega\kelvin} (Eq. \ref{eq:code:lambda_t}). As density stays almost constant after $t=$ \SI{175}{\second}, the radiative losses become less efficient in cooling the plasma. Thus, the temperature continuously increases after that time in the manner of a thermal instability, until conduction becomes sufficiently strong in the corona to compensate for the constant atmospheric heating. Given enough time ($\sim$ \SI{e4}{\second}), the final temperature and density reach those of the newly formed final hot loop in equilibrium, which is also the initial state of m$_{23}$. Regarding m$_{23}$, the peak in temperature is also reached in about \SI{10}{\second}, followed by a slower cooling. The density, however, remains the same overall before $t = $ \SI{120}{\second}. At this point, we can see some oscillations in the profile. The density at $t=$ \SI{300}{\second} is slightly above the one at $t=$ \SI{100}{\second}, because of evaporation following the impulsive heating (from $\log{n}$ = 1.9 to 2.4). Given enough time ($\sim$ \SI{e4}{\second}), the hot loop eventually returns to its initial hot state in equilibrium. These results are important for understanding the light curves behavior in the next section. 

\subsection{Light curves}
\label{sec:results:lc}

In this section, we present the synthetic light curves of HRIEUV, four EUV channels of AIA, and five lines measured by SPICE. They are computed from the simulation results of the groups I (section \ref{sec:results:lc:changing_L}) and III models (\ref{sec:results:lc:changing_H}). The response and contribution functions of all channels and lines (Fig. \ref{fig:annex:gofnt}) are computed with CHIANTI \citep{Dere_Chianti}, version 10.1 \citep{Del_zanna_chianti_2021,Dere_2023}. We assumed the coronal abundance estimated by \cite{Asplund_2021} and the ionization equilibrium recommended by CHIANTI.

\subsubsection{Light curves computed from group I models}
\label{sec:results:lc:changing_L}

\renewcommand{\arraystretch}{1.3}
\begin{table*}
\caption{Parameters of the group I models.}
\flushleft          
\begin{tabular}{c c c c | c c c | c | c c c c c}   
\multicolumn{13}{c}{Group I} \\
\hline       
\hline       
\multicolumn{4}{c|}{Initial cool loop} & \multicolumn{3}{c|}{Initial hot loop} &
\multicolumn{1}{c|}{} & \multicolumn{5}{c}{Heating} \\
\hline       
  Name & $T_\mathrm{iCL}$ & $n_\mathrm{iCL}$ &  $P_0$  & Name & $T_\mathrm{iHL}$ & $n_\mathrm{iHL}$ & $L$ &  $H_0$ & $A_{\mathrm{max}}$ & $\sigma_\mathrm{ev}$ & $s_\mathrm{ev} - L_\mathrm{mid}$ & iCL $\rightarrow$ fHL
\\
\hline
$m_1$ & 4.4 & 10.1 & 0.1 & \textit{N/a} & \textit{N/a} & \textit{N/a} & 0.5 & 0.023 & 0.2 & 1.0 & 0 & No  \\
$m_2$ & 4.5 & 10.0 & & \textit{N/a} & \textit{N/a} & \textit{N/a} & 1.0 &  & & 2.0 & & No  \\
$m_3$ & 4.6 & 9.8 & &  $m_{22}$ & 5.7 & 9.2 & 2.0 & & & 4.0 & & Yes  \\
$m_4$ & 4.8 & 9.5 & & $m_{23}$ & 5.8 & 9.3 & 3.0 & & & 6.0 & & Yes  \\
$m_5$ & 4.9 & 9.4 & & $m_{24}$ & 5.9 & 9.3 & 4.0 & & & 8.0 & & Yes  \\
$m_6$ & 4.9 & 9.3  & & $m_{25}$ & 5.9 & 9.3 & 5.0 & & & 10.0 & & Yes  \\
$m_7$  & 5.0 & 9.3 & & $m_{26}$ & 5.9 & 9.3 & 5.5 & & & 11.0 & & Yes  \\

\end{tabular}
\tablefoot{Initial parameters of the group I models, with an increasing length $L$ in the central part of the loop. The initial electron temperatures ($\log{\mathrm{K}}$) and densities ($\log{\mathrm{cm^{-3}}}$) at the apex are respectively indicated for an initial cool ($T_\mathrm{iCL}$, $n_\mathrm{iCL}$) and hot ($T_\mathrm{iHL}$, $n_\mathrm{iHL}$) loops in equilibrium. For the cool loop in equilibrium, the pressure at the top of the vertical chromospheric legs is given by $P_0$ (\SI{}{\dyne\per\centi\meter\tothe{2}}). $H_0$ (\SI{}{\erg\per\centi\meter\tothe{3}\per\second}) is the uniform and constant term of the atmospheric heating $H$. The impulsive heating has an maximal amplitude $A_\mathrm{max}$ (\SI{}{\erg\per\centi\meter\tothe{3}\per\second}), a spatial standard deviation $\sigma_\mathrm{ev}$ (\SI{}{\mega\meter}), and a central position $s_\mathrm{ev}$ (\SI{}{\mega\meter}). The latter is given with respect to the center of the loop $L_\mathrm{mid} = L_\mathrm{c} + \frac{L}{2}$ (\SI{}{\mega\meter}). The column “iCL $\rightarrow$ fHL” indicates whether the initial cool loop has evolved into a final hot loop (marked as fHL) following the impulsive heating. Constant terms over a column are indicated by blanks, and \textit{N/a} means that no hot loop in equilibrium could be built with the given parameters.} 
\label{table:code:groupI}      
\end{table*}

\begin{figure*}
    \centering
    \includegraphics{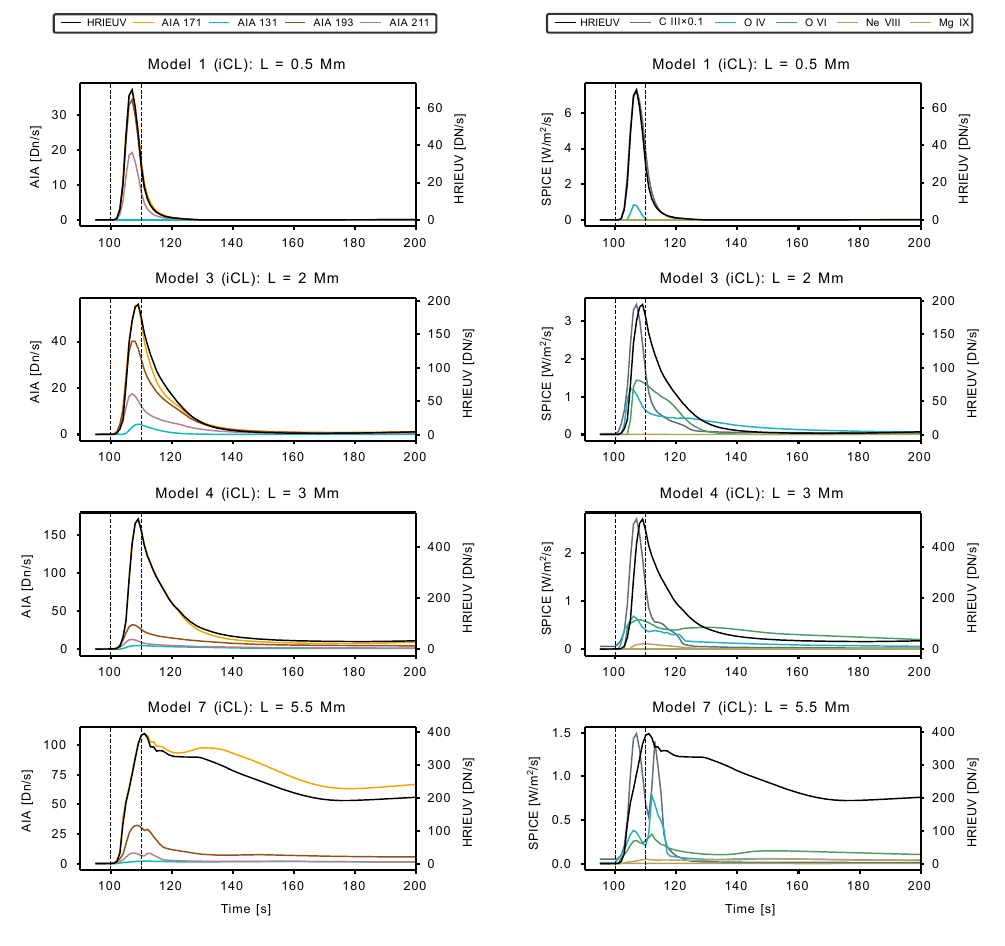}
    \caption{Light curves for the simulations of four group I models, with loops in an initial cool state (Table \ref{table:code:groupI}). In this case, the length $L$ of the loop's central part increases from \SI{0.5}{\mega\meter} to \SI{5.5}{\mega\meter}. The intensities correspond to four AIA channels (left column), five SPICE lines (right column), and HRIEUV (on the right axis of all subfigures). For better visibility, the intensity of the \ion{C}{iii} line is divided by 10. Results are presented between $t=$ \SI{90}{} and \SI{200}{\second}, with the start ($t=$ \SI{100}{\second}) and the end times ($t=$ \SI{110}{\second}) of the impulsive heating indicated by two vertical dotted lines.}
    \label{fig:results:lc:CL_HL_groupI_these}
\end{figure*}

\begin{figure*}
    \centering
    \includegraphics{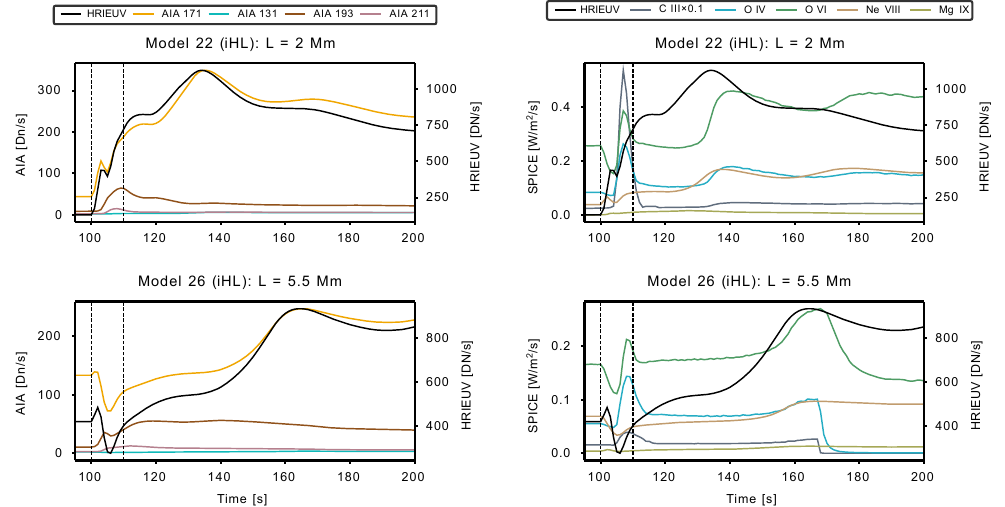}
    \caption{Light curves for the simulations of two group I models, with loops in an initial hot state (Table \ref{table:code:groupI}). This figure is similar to Fig. \ref{fig:results:lc:CL_HL_groupI_these}.}
    \label{fig:results:lc:HL_HL_groupI_these}
\end{figure*}

We present here the light curves results for the group I models (Table \ref{table:code:groupI}). As a reminder, loops in this group have an increasing length $L$ from \SI{0.5}{} to \SI{5.5}{\mega\meter}. Apart from m$_1$ and m$_2$, all cool loops evolve into a hot loop after impulsive heating. 

Figure \ref{fig:results:lc:CL_HL_groupI_these} shows the results for models with loops in an initial cool state. For the models m$_1$ to m$_4$, all light curves reach their intensity peak less than \SI{10}{\second} after the start of the impulsive heating. The peak is almost co-temporal among most light curves, with less than a \SI{5}{\second} delay. This delay does not vary much with the loop length. Bright emission depends on both temperature and density. The temperature must be well matched with the spectral line or observing channel, and the density must be high to have a large emission measure. As shown in Figure \ref{fig:results:m4_m23_n_t_apex}, density is highest in the initial equilibrium state and for the first \SI{25}{\second} after the start of the nanoflare. Temperature increases quickly during the nanoflare, and the different emissions turn on in rapid succession. Thus, the rise phases of the light curves are nearly co-temporal. After \SI{25}{\second}, the density decreases as material flows down the legs and the loop transitions to a hot equilibrium. The light curves decay in a similar manner from a combination of the decreasing density and decreasing temperature. This is further discussed in Section \ref{sec:results:flows:cl}. Regarding m$_7$, we notice two successive intensity peaks in the \ion{C}{iii}, \ion{O}{iv}, and \ion{O}{vi} lines. This is also caused by density fluctuations due to flows.

Figure \ref{fig:results:lc:HL_HL_groupI_these} shows the results for models with loops in an initial hot state. The light curves of hot loops are complex and dramatically different from those of cool loops. They have two components -- an initial impulsive component followed by a gradual component. The relative strength of the components varies with the temperature of the channel or line. The gradual component is generally stronger, but the impulsive component has comparable brightness in cooler emissions. For instance, the impulsive peak of m$_{22}$ at $t=$\SI{105}{\second} is strong in the SPICE light curves, but almost nonexistent in the HRIEUV and AIA light curves. The impulsive component is a response of the lower atmosphere to a sudden and intense downward thermal conduction flux brought on by the nanoflare \citep[][]{Qiu_2013}. The gradual component is a combination of two distinct effects of varying importance: the cooling through conduction and radiation of the coronal part of the loop, and density variations as the loop is filled with evaporating material and subsequently drained. Note that, in contrast to cool loops, the emission is faint along most of the loop during the heating phase because equilibrium hot loops have lower density than equilibrium cool loops (section \ref{sec:results:temperature_loop}).

Regarding m$_{22}$ and m$_{26}$, the gradual component mainly reflects the slow cooling of the coronal part of the loop from $\log{T} > 6.3$, through conduction and radiation. As the corona is relatively isothermal, we observe significant delays between the intensity peaks of the AIA and HRIEUV light curves, as the temperature of the corona passes through the maximum of the response function for the hotter (AIA 211, 193) to the cooler channels (HRIEUV, AIA 171). For example, the AIA 193 and 171 intensity peaks are separated by \SI{30}{\second} for m$_{22}$ (L = \SI{2}{\mega\meter}) and \SI{50}{\second} for m$_{26}$ (L = \SI{5.5}{\mega\meter}). With regards to SPICE (Fig. \ref{fig:results:lc:HL_HL_groupI_these}, right column), the behaviors of the light curves are complex, with multiple peaks and oscillations. This behavior can be explained by a combination of temperature variation and the fluctuation of density along the loop because of plasma flows. These flows are further described in section \ref{sec:results:flows:hl}. Because the flows originate mainly from plasma at TR temperatures, they have more impact on the behavior of the SPICE light curves (mostly sensitive to TR emission) compared to those of AIA and HRIEUV (mostly sensitive to high TR to coronal emission).

As a reminder, the initial hot loops in Fig. \ref{fig:results:lc:HL_HL_groupI_these} are obtained from the final states of the simulations shown in Fig.\ref{fig:results:lc:CL_HL_groupI_these}. For instance, the initial hot state of m$_{23}$ is given by the final hot state of m$_4$ (Table \ref{table:code:groupI}). However, it takes more than \SI{600}{\second} for the hot loops to reach equilibrium, to which we add $\sim$ \SI{e4}{\second} of relaxation time (Section \ref{sec:code:cl_hl}). This is why the light curves at $t=$ \SI{200}{\second} in Fig. \ref{fig:results:lc:CL_HL_groupI_these} tend to (but are not equal to) those at $t=$ \SI{0}{\second} in Fig. \ref{fig:results:lc:HL_HL_groupI_these}.

For nine of the events, \cite{Dolliou_2024} measured the light curves of all of the channels and lines displayed in Fig. \ref{fig:results:lc:CL_HL_groupI_these} and \ref{fig:results:lc:HL_HL_groupI_these}. Fig.6 from their paper, which shows the light curves for three events, is reproduced in Fig.\ref{fig:annex:fig6}. For most events they studied, the intensity peaks between all channels and lines had short delays below their respective cadence (\SI{6}{\second} for AIA and \SI{25}{\second} for SPICE). In our work, the delays observed for models with loops in an initial cool state (Fig. \ref{fig:results:lc:CL_HL_groupI_these}) are below these limits, so the results are consistent with observations. On the other hand, for models with loops in an initial hot state (Fig. \ref{fig:results:lc:HL_HL_groupI_these}), the delays are too large to be consistent with observations.

We also focus on the relative maximal intensity reached by each EUV channel and line. For all models, HRIEUV reaches higher intensity values than other EUV channels, followed by AIA 171, 193, 211, and 131. As for SPICE, the intensity peak of the \ion{C}{iii} line is higher than those of \ion{O}{vi}, \ion{O}{iv}, and \ion{Ne}{viii}. The \ion{Mg}{ix} intensity is negligible compared to those of the other lines. This is consistent with observational studies, in which no emission in \ion{Mg}{ix} was ever reported \citep[][]{Dolliou_2024,Huang_2023}.

Lastly, we notice that the HRIEUV intensity can reach values of the same magnitude for models with cool loops (up to \SI{150}{\dn\per\second}, m$_4$) and hot loops (up to \SI{300}{\dn\per\second}, m$_{23}$), despite some of them not reaching coronal temperatures. This is due to the broad response function of HRIEUV, which covers temperatures from the TR to the corona (Fig.\ref{fig:annex:gofnt}). This highlights the fact that events detected by HRIEUV do not necessarily reach coronal temperatures \citep[see also][]{Tiwari2022}.

\subsubsection{Light curves computed from group III models}
\label{sec:results:lc:changing_H}

\begin{table*}
\caption{Parameters of the group III models.}
\flushleft          
\begin{tabular}{c c c c | c c c | c | c c c c c}   
\multicolumn{13}{c}{Group III} \\
\hline       
\hline       
\multicolumn{4}{c|}{Initial cool loop} & \multicolumn{3}{c|}{Initial hot loop} &
\multicolumn{1}{c|}{} & \multicolumn{5}{c}{Heating} \\
\hline       
Name & $T_\mathrm{iCL}$ & $n_\mathrm{iCL}$ &  $P_0$  & Name & $T_\mathrm{iHL}$ & $n_\mathrm{iHL}$ & $L$ &  $H_0$ & $A_{\mathrm{max}}$ & $\sigma_\mathrm{ev}$ & $s_\mathrm{ev} - L_\mathrm{mid}$ & iCL $\rightarrow$ fHL
\\
\hline
$m_{14}$ & 4.8 & 9.5 & 0.1 & $m_{33}$ & 5.8 & 9.3 & 3.0  & 0.023 & 0.05 & 6.0 & 0 & No \\
$m_{15}$ &  & & &  $m_{34}$ & &  &  &  & 0.1 & & & Yes  \\
$m_{4}$ &  & & &  $m_{23}$ & &  &  &  & 0.2 & & & Yes  \\
$m_{16}$ &  & & &  $m_{35}$ & &  &  &  & 0.5 & & & Yes  \\

\end{tabular}
\tablefoot{Initial parameters of the group III models, with variation in the impulsive heating maximal amplitude $A_{\mathrm{max}}$. The parameters are the same as for Table \ref{table:code:groupI}.} 
\label{table:code:groupIII}      
\end{table*}

\begin{figure*}
    \centering
    \includegraphics{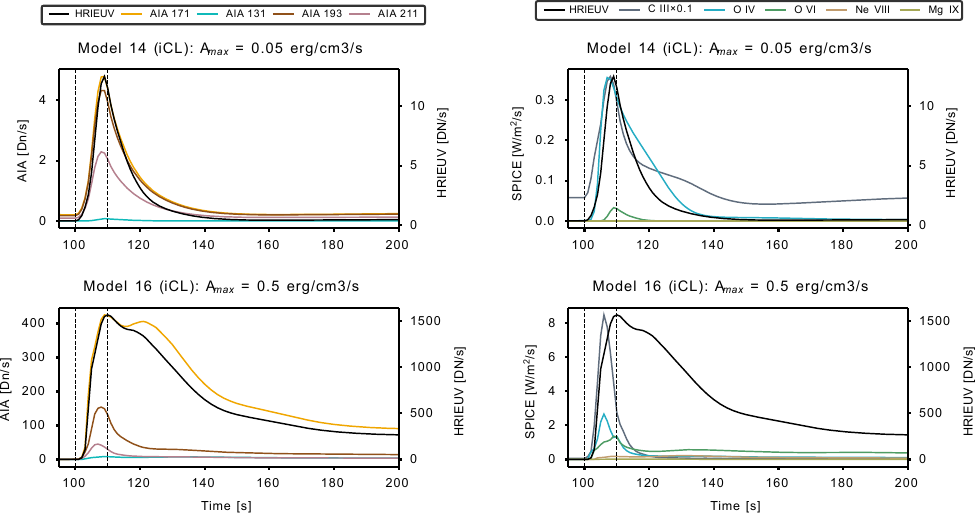}
    \caption{Light curves for the simulations of two group III models, with loops in an initial cool state (Table \ref{table:code:groupIII}). This figure is similar to Fig. \ref{fig:results:lc:CL_HL_groupI_these}.}
    \label{fig:results:lc:CL_HL_groupeIII_these}
\end{figure*}

\begin{figure*}
    \centering
    \includegraphics{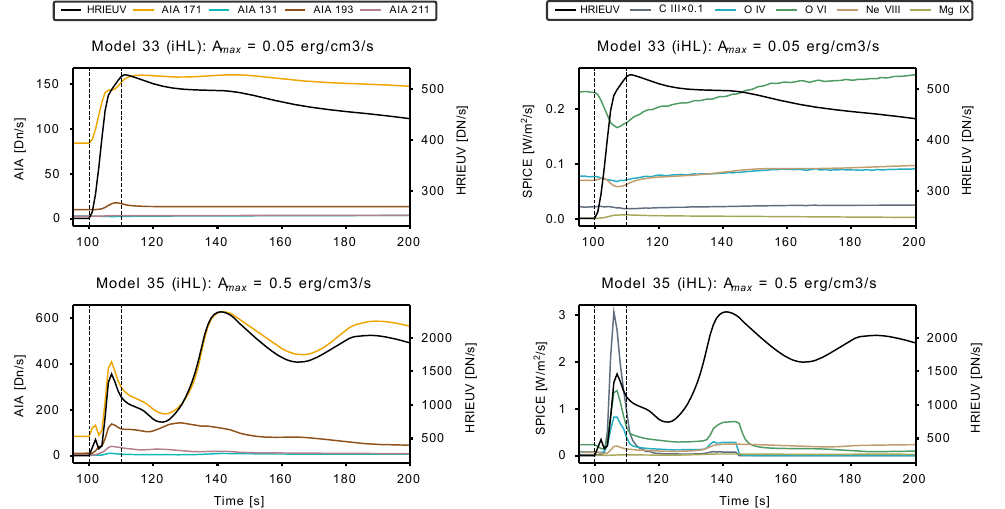}
    \caption{Light curves for the simulations of two group III models, with loops in an initial hot state (Table \ref{table:code:groupIII}). This figure is similar to Fig. \ref{fig:results:lc:HL_HL_groupI_these}.}
    \label{fig:results:lc:HL_HL_groupeIII_these}
\end{figure*}
Here, we present the light curves computed from the group III models, with impulsive heating increasingly stronger (Table \ref{table:code:groupIII}). Figure \ref{fig:results:lc:CL_HL_groupeIII_these} shows the light curves computed from the simulations with loops in an initial cool state. The results are similar to those obtained in Fig. \ref{fig:results:lc:CL_HL_groupI_these}. Indeed, in both cases, we observe co-temporal intensity peaks between all light curves, less than \SI{10}{\second} after the start of the impulsive heating. Furthermore, the intensity of HRIEUV increases above the background for less than \SI{30}{\second} after impulsive heating in model m$_{14}$ (as in model m$_3$), and for more than \SI{100}{\second} in m$_{16}$ (as in m$_7$). The relative maximal intensities reached by all channels and lines are also similar to those from group I models. 

Figure \ref{fig:results:lc:HL_HL_groupeIII_these} shows the light curves of models with loops in an initial hot state. The strength of the impulsive component relative to the gradual component increases with increasing nanoflare energy. This is expected, as higher energies produce higher peak temperatures and therefore a much stronger downward conductive flux to the footpoint ($\propto T^{5/2}$). For the model m$_{35}$ ($A_{\mathrm{max}} = $ \SI{0.5}{\erg\per\centi\meter\tothe{3}\per\second}), we observe two co-temporal intensity peaks in HRIEUV, AIA 171, and most SPICE lines. Some aspects of this model are consistent with the observations. 

In observations, successive peaks are frequently observed for events detected in HRIEUV \citep[for instance, Fig. 6a,c in][, reproduced in Fig. \ref{fig:annex:fig6}]{Dolliou_2024}. These successive peaks can have a similar origin to those seen in m$_{35}$, but it is not the only possible explanation. Indeed, successive peaks can also be the signature of multiple impulsive energy releases. It is also worth noting that light curves in \cite{Dolliou_2024} are also obtained by averaging the intensities over a spatial region around the events. Thus, distinct and unresolved magnetic structures can also be responsible for the successive intensity peaks. Furthermore, contrary to observations, not all light curves of m$_{35}$ are co-temporal with each other. In fact, we still measure delays of about \SI{15}{\second} between the AIA 171 and 193 intensity peaks. Consequently, the observational results are reproduced only in part.

\subsection{Impact of high density effects and plasma flows on the light curve behavior}
\label{sec:results:flows}

We now aim to explain the formation processes of the co-temporal intensity peaks observed between multiple HRIEUV, AIA, and SPICE light curves, which are consistent with observations (section \ref{sec:results:lc}). In Section \ref{sec:results:flows:cl}, we use m$_4$ as an example to discuss their origin in models with loops in an initial cool state that evolve into hot loops. Section \ref{sec:results:flows:hl} presents the only case of a model with a loop in an initial hot state (m$_{35}$), where two co-temporal intensity peaks appear successively.

\subsubsection{Cool loop evolving into a hot loop}
\label{sec:results:flows:cl}

\begin{figure*}
    \centering
    \includegraphics{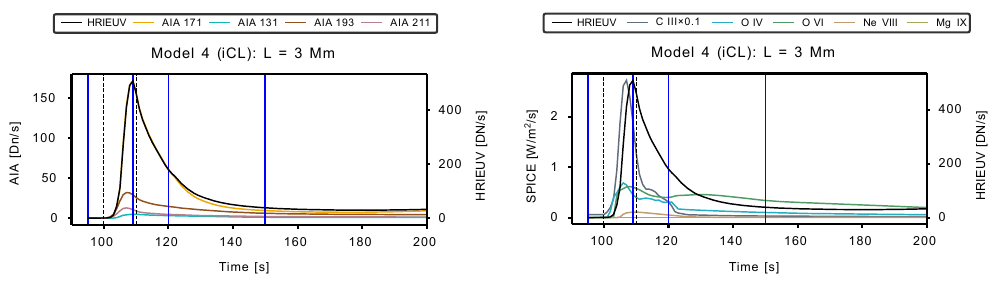}
    \includegraphics{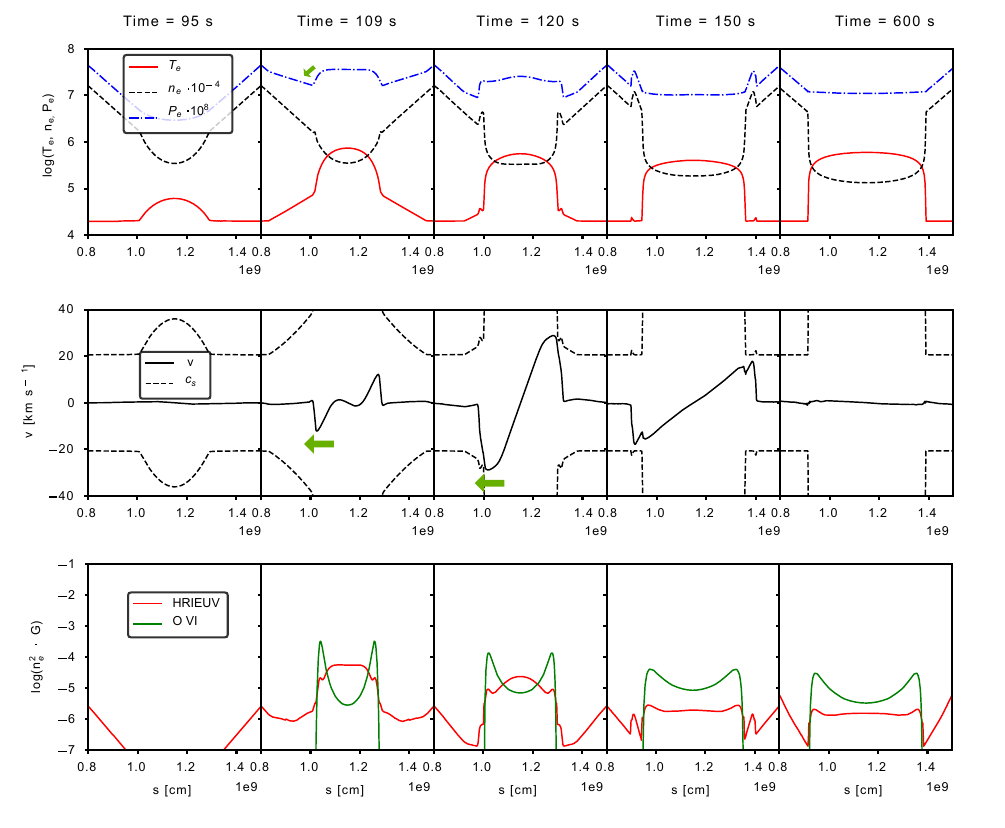}
    \caption{Results for model m$_4$, with an initial cool loop. First row from the top: Light curves of HRIEUV, the AIA channels (left), and the SPICE lines (right) between $t=$ \SI{95}{\second} and \SI{200}{\second}. The vertical dotted black lines show the beginning and end of the impulsive heating at \SI{100}{\second} and \SI{110}{\second}. The vertical blue lines indicate the times studied in the panels below. These panels are the same as in Fig. \ref{fig:results:lc:HL_HL_groupI_these}. Second row: Electron temperature, density, and pressure profiles around the loop's central part. Green arrows show the pressure gradients that are responsible for the plasma flows. Third row: Velocity $v$ and sound speed $c_{\mathrm{s}}$ profiles. Green arrows indicate the direction of the flows. Fourth row: HRIEUV (in \SI{}{\dn\per\centi\meter\per\second}) and \ion{O}{vi} (in \SI{}{\milli\watt\per\meter\tothe{3}\per\second}) emissivity profiles along the loop. The flows are discussed in detail in section \ref{sec:results:flows:cl}.}

    \label{fig:results:CL_HL_m4_lc_ntv}
\end{figure*}

Figure \ref{fig:results:CL_HL_m4_lc_ntv} (1st row) displays the light curves computed for m$_4$ (Table \ref{table:code:groupI}), with a loop in an initial cool state. Specific times (blue vertical lines) are selected for further investigations. The rows 2 to 4 display various parameters at these selected times: the electron temperature, density, and pressure profiles ($2^{\mathrm{nd}}$ row); the plasma velocity and the sound speed profiles ($3^{\mathrm{th}}$ row); the HRIEUV intensity and the \ion{O}{vi} emission profiles (4$^{\mathrm{th}}$ row). The temperature and density profiles at $t=$ \SI{95}{\second} (2$^{\mathrm{nd}}$ row) are those of the initial cool loop in equilibrium, while the profiles at $t=$ \SI{600}{\second} are those of the final hot loop in equilibrium.

We now explain the physical origin of the co-temporal intensity peaks between most of the light curves at $t = $ \SI{109}{\second} (Fig. \ref{fig:results:CL_HL_m4_lc_ntv}, $1^{\mathrm{st}}$ row). In the first part, between $t=$ \SI{95}{\second} and $t=$ \SI{109}{\second}, the temperature increases from $\log{T_\mathrm{apex}} = 4.8$ to $5.9$, while the density stays constant around $\log{n_{\mathrm{apex}}} = 9.6$ (2$^{\mathrm{nd}}$ row). Due to the high density along the loop and the sudden increase in temperature to TR values, the intensity of the channels sensitive to the emission of plasma below $\log{T} = 6.0$ will significantly increase co-temporally during this period (1$^{\mathrm{st}}$ row). In particular, both HRIEUV and the \ion{O}{vi} lines are sensitive to emission at TR temperatures (Fig. \ref{fig:annex:gofnt}), so their emission increases everywhere along the loop between $t=$ \SI{100}{\second} and $t=$ \SI{109}{\second} (4$^{\mathrm{th}}$ row). In the second part, between $t=$ \SI{109}{\second} and $t=$ \SI{140}{\second}, the cool loop evolves into a hot state. As part of this transition, the density decreases in the central part of the loop from $\log{n_{\mathrm{apex}}} = 9.6$ to $9.1$, while the temperature decreases from $\log{T_{\mathrm{apex}}} = 5.9$ to $5.5$. Because the temperature remains in the TR range, this sharp density decrease is responsible for the co-temporal decrease in the intensity of all channels and lines between $t=$ \SI{109}{\second} and $t=$ \SI{140}{\second} (1$^{\mathrm{st}}$ row). Indeed, the intensity of a channel or a line depends on the square of the plasma density (Eq. \ref{eq:annex:I}).

The decrease in density is caused by flows that eject the plasma from the loop center to the chromosphere. They are displayed by the green arrows (3$^{\mathrm{rd}}$ row). Their velocity can reach \SI{23}{\kilo\meter\per\second}. These velocities remain subsonic, despite being at some point almost equal to the sound speed, at t=\SI{120}{\second}. These flows are caused by a rupture in the hydrostatic equilibrium due to the impulsive heating. Although the Gaussian spatial profile of the heating is very broad ($\sigma_{\mathrm{ev}}$ = $2L$), the heating amplitude is so large that it produces sizable temperature and pressure gradients directed away from the loop apex. This is clearly seen at t = \SI{109}{\second} in Figure \ref{fig:results:CL_HL_m4_lc_ntv} (2$^{\mathrm{nd}}$ row). The resulting evacuation of plasma from the loop accommodates the eventual hot equilibrium, which has a much lower density than the original cool equilibrium.

\subsubsection{Hot loop submitted to a strong impulsive heating}
\label{sec:results:flows:hl}

\begin{figure*}
    \centering
    \includegraphics{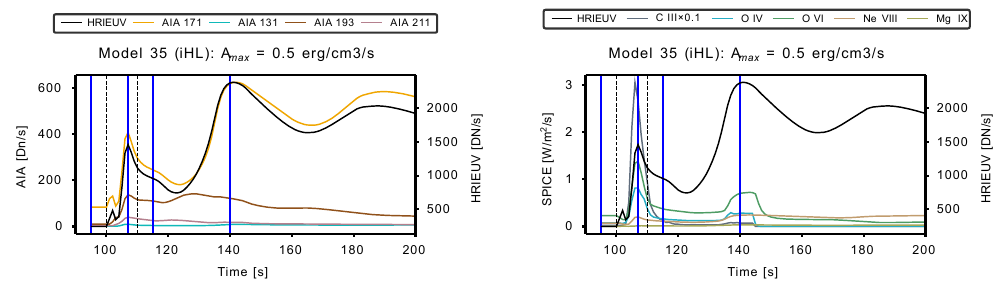}
    \includegraphics{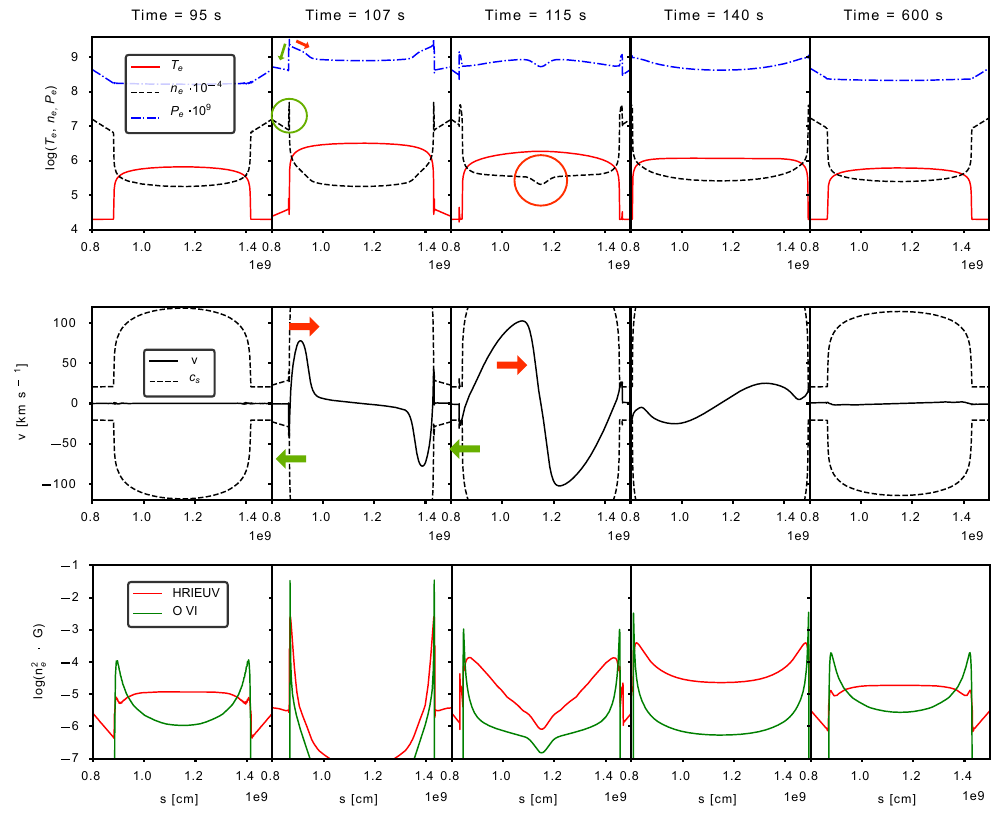}
    \caption{Same as for Fig.\ref{fig:results:CL_HL_m4_lc_ntv}, but for the model m$_{35}$, which consists of an initial hot loop submitted to an impulsive heating with a large maximal amplitude of $A_{\mathrm{max}} = $ \SI{0.5}{\erg\per\centi\meter\tothe{3}\per\second}. Green and red arrows highlight the flows responsible for the density fluctuations indicated by the respective green and red circles. These flows are discussed in detail in section \ref{sec:results:flows:hl}.   }

    \label{fig:results:HL_HL_m35_lc_ntv}
\end{figure*}

In this section, we use Fig. \ref{fig:results:HL_HL_m35_lc_ntv} to describe the origin of the two successive co-temporal intensity peaks between most of the light curves for m$_{35}$. These peaks can be seen at $t=$ \SI{107}{\second} and \SI{143}{\second} (Fig. \ref{fig:results:HL_HL_m35_lc_ntv}, 1$^{\mathrm{st}}$ row). We distinguish two types of flows, highlighted by the green and red arrows in the velocity profiles (3$^{\mathrm{rd}}$ row). In the following, they will be referred to as the “downflows” (green arrow) and the “upflows” (red arrow). 

Between $t=$ \SI{95}{\second} and \SI{107}{\second}, the temperature in the coronal part of the loop increases from $\log{T_{\mathrm{apex}}} = 5.8$ to $6.5$ after impulsive heating. An intense downward conduction flux forms from the elevated coronal temperatures and heats the top of the chromosphere. The localized high pressure drives evaporating upflows along the loop (3$^{\mathrm{rd}}$ row, red arrow) and downflows (green arrow). The downflows evolve into propagating shock fronts at $t =$ \SI{107}{\second}, as their velocity (up to \SI{40}{\kilo\meter\per\second}) exceeds the local speed of sound in the chromosphere ($\approx$ \SI{30}{\kilo\meter\per\second}). This results in a dense front forming in the lower TR (2$^{\mathrm{nd}}$ row, green circle). Consequently, co-temporal intensity peaks form in multiple channels and lines at $t=$ \SI{107}{\second} (1$^{\mathrm{st}}$ row). The emissions mainly come from the location of the dense front, as seen in the HRIEUV and OVI profiles (4$^{\mathrm{th}}$ row). This highly compressed plasma cools down fast, through radiation and conduction, while propagating into the chromosphere. As a result, the light curves of all channels will decrease in quick succession, appearing co-temporal.

The evaporating upflows are responsible for density fluctuations throughout the loop, for instance, at $t = $ \SI{115}{\second} (2$^{\mathrm{nd}}$ row, red circle). Their velocity can reach up to \SI{100}{\kilo\meter\per\second}, which is below the sound speed in the coronal part of the loop. The evaporative upflows rebound when they collide at the top of the loop. When the resulting downflows hit the TR at $t=$ \SI{143}{\second}, we observe co-temporal intensity peaks in HRIEUV, AIA 171, and most of the SPICE lines. Then, some of the energy continues propagating downward, and some of the energy is reflected back into the loop. The reflection is stronger in hot loops than in cool loops because of the steep density gradient. We then observe other intensity peaks in the HRIEUV and in the AIA 171 light curves, as the reflected flows reach the TR again (1$^{\mathrm{st}}$ row, at $t=$ \SI{190}{\second}).  

\section{Discussion}
\label{sec:discussion}

The aim of our work was to propose a physical explanation for the small EUV brightenings detected by HRIEUV. Assuming they originate from unresolved short loops, we simulated the evolution of impulsively heated short loops in agreement with the events' properties (apparent length, lifetime, height). We used two models for the initial equilibrium conditions of the loops: cool ($T < $ \SI{e5}{\kelvin}) and hot loops ($T > $ \SI{e5}{\kelvin}). We performed a parametric analysis over the loops (groups I, II, and IV) and impulsive heating properties (groups III and V). The temperature and density outputs of the simulations were used to compute the synthetic light curves for HRIEUV, four EUV channels of AIA, and five lines measured by SPICE. We assumed the short loops not to be resolved during observations. Therefore, the light curves are computed by averaging the intensity of all LOS covering the loop. The parameters of the models are then constrained by comparing the behavior of the synthetic light curves with those obtained from observations. In particular, we wanted to reproduce the co-temporal intensity peaks seen for most events in the HRIEUV, the SPICE, and the AIA light curves. \citep[][]{Dolliou_2023,Dolliou_2024}. 

In the following sections, we discuss our results in three main points. In section \ref{sec:discussion:A}, we explain why impulsively heated cool loops are good candidates to explain the origin of events detected in the HRIEUV sequences. In section \ref{sec:discussion:B}, we also discuss the alternative scenario of a hot loop subjected to a strong impulsive heating. Finally, in section \ref{sec:discussion:C}, we highlight how high plasma density and flows may affect the light curves behavior. 

\subsection{Cool loops are good candidates to explain events detected with HRIEUV}
\label{sec:discussion:A}

The models that use cool loops as the initial equilibrium state have light curve behaviors in agreement with observations \citep[][]{Dolliou_2024,Huang_2023} and the events' lifetime \citep[10 to \SI{200}{\second}][]{Berghmans2021}. Indeed, three key observational results are reproduced: (1) the light curves showed a co-temporal peak (Fig. \ref{fig:results:lc:CL_HL_groupI_these} and \ref{fig:results:lc:CL_HL_groupeIII_these}); (2) after the impulsive heating, the HRIEUV light curve peaks and then decreases in less than \SI{100}{\second}; and (3) the \ion{Mg}{ix} intensity peak was negligible compared to the other emission lines measured by SPICE.

This compatibility with the observations was found for a wide range of parameters affecting the geometry of the loop (groups I, II, IV) and the heating (group III). Therefore, we suggest that impulsively heated cool loops are good candidates to explain the origin of events detected by HRIEUV.

There are two exceptions to this claim. First is the m$_{16}$ model of a cool loop subjected to strong impulsive heating. Due to the high density ($\log{n} = 9.5$) and the coronal temperature ($\log{T} = 6.2$) reached at the apex (Fig. \ref{fig:results:fig1_tmin_tmax}), we expect this loop to be (barely) detected in the \ion{Mg}{ix} \SI{706.02}{\angstrom} line. The observations analyzed in \cite{Dolliou_2024} set the limits of the \ion{Mg}{ix} \SI{706.02}{\angstrom} line detection to $\log{n}= 9.36$ at $\log{T} = 6.0$. Therefore, the model m$_{16}$ might not be compatible with the fact that no \ion{Mg}{IX} \SI{706.02}{\angstrom} emission associated with events has ever been observed.

The second exception is the m$_{20}$ model of a cool loop subjected to narrow impulsive heating localized on the left leg of the loop's central part (Section \ref{annex:sec:groupV_results}). In that case, contrary to observations, there are delays up to \SI{30}{\second} between the intensity peaks of the \ion{O}{vi}, \ion{O}{iv}, and \ion{C}{iii} light curves (Fig. \ref{fig:annex:group5_light_curves}). Therefore, under the assumption of an unresolved loop, this is an indication that the impulsive heating might be centered near the apex of the loop. We note that similar conclusions were drawn by \cite{Zhukov2021}, assuming a semicircular shape for the loops.

We note that the hypothesis of a cool loop evolving into a hot loop can only explain the origin of isolated EUV brightenings, which do not repeat on the same loop. To explain the origin of the latter, one should consider the possibility of an initial hot loop, as discussed in the next section.

\subsection{Hot loops are not consistent with observations, except for strong impulsive heating}
\label{sec:discussion:B}

Impulsively heated hot loops show significant delays of up to more than \SI{30}{\second} between the intensity peaks of the AIA 171 and 193 light curves (Sections \ref{sec:results:lc:changing_L} and \ref{sec:results:lc:changing_H}). As such, most models with loops in an initial hot state are not consistent with observations. There is one exception, when the hot loop is subjected to strong impulsive heating (m$_{35}$). In that case, the following observational properties are reproduced: (1) the HRIEUV, the AIA 171, and the SPICE light curves have two successive co-temporal intensity peaks with \SI{35}{\second} of delay (Section \ref{sec:results:lc:changing_H}); (2) we expect the \ion{Mg}{ix} \SI{706.02}{\angstrom} line to remain undetected even if the loop reaches coronal temperature ($\log{T} = 6.5$), because its density ($\log{n} = 9.25$, Fig.\ref{fig:results:fig1_tmin_tmax}) remains below the confidence level to detect the line \citep[$\log{n} = 9.36$,][]{Dolliou_2024}. However, not all results of m$_{35}$ are in agreement with observations, as delays up to $\approx$ \SI{15}{\second} still exist between the AIA 171 and AIA 193 intensity peaks. Nevertheless, we suggest that hot loops subjected to strong impulsive heating are good candidates to explain the origin of some events, including those showing two successive intensity peaks \citep[Fig 6a,c][]{Dolliou_2024} and those that appear in short loops already visible in HRIEUV prior to them \citep[Fig.5a][]{Dolliou_2024}. Indeed, we can expect the initial hot loop at TR temperature to be visible in HRIEUV, considering the response function of the instrument (Fig. \ref{fig:annex:gofnt}).

The m$_{35}$ model is an exception among models with loops in an initial hot state. Downward-propagating shock fronts and evaporating upflows (up to \SI{100}{\kilo\meter\per\second}) form at the top of the chromosphere due to strong conductive flux. The latter is thus required to create co-temporal intensity peaks between the light curves that are consistent with observations (Sect. \ref{sec:results:flows:hl}). It can be induced by strong impulsive heating (m$_{35}$) or a lower pressure in the loop (m$_{30}$, Fig.\ref{fig:annex:group2_light_curves}). There are some published arguments supporting the claim that events are associated with strong impulsive heating. First, they are mainly distributed around the chromospheric network \citep[][]{Berghmans2021}, where the radial component of the photospheric Poynting flux tends to be larger ($\left|S_\mathrm{z}\right| \sim$ \SI{e8}{\erg\per\centi\meter\tothe{2}\per\second}) than the rest of the QS \citep[$ \left|S_\mathrm{z}\right|\sim$ \SI{e7}{\erg\per\centi\meter\tothe{2}\per\second},][]{Tilipman_2023}. While most of the energy is released and the chromosphere, and taking into account the expansion of the magnetic flux, we can still hypothesize that the energy release below \SI{5000}{\kilo\meter} above the photosphere is higher near the chromospheric network compared to the rest of the QS. Second, \cite{Nelson_2024} found that the events are mainly located above strong photospheric magnetic field regions (around 80.1\% of them). A stronger field can then result in larger energy release compared to the rest of the QS. Consequently, it is reasonable to suggest that at least part of the events might correspond to the m$_{35}$ case.

Despite being subject to potentially strong impulsive heating, there are also arguments claiming that the events' emission is, for the most part, not coronal. \cite{Dolliou_2024} estimated the DEM of one event by applying an inversion algorithm on the lines intensities measured with EIS. They find that the event's emission mainly comes from plasma at TR temperatures ($T\sim$ \SI{e5}{\kelvin}). This suggests that this specific event is not explained by the m$_{35}$ model. If that were the case, the event should be associated with coronal emission visible in EIS given its typical coronal temperature (up to $\log{T} = 6.5$) and density ($\log{n} = 9.25$, Fig. \ref{fig:results:fig1_tmin_tmax}) at the loop apex. Nevertheless, the last statement is subject to the instrumental limitations of EIS, including its spatial resolution ($3^{\prime \prime}$\footnote{\url{https://sohoftp.nascom.nasa.gov/solarsoft/hinode/eis/doc/eis_notes/08_COMA/eis_swnote_08.pdf}, consulted on 2024 April 16.}). Assuming the events have a small filling factor, the emission from plasma at coronal temperature might remain below noise level.

\subsection{High density and plasma flows at the origin of co-temporal intensity peaks}
\label{sec:discussion:C}

As a reminder, \cite{Dolliou_2023} measured the delays between the intensity peaks of pairs of AIA channels associated with the QS events detected with HRIEUV. For nine pairs, they showed that most events were characterized by short delays below the AIA cadence (\SI{12}{\second}), with some variation around the mean value below \SI{1}{\minute}. They suggested two possible interpretations for the short delays of the main event population: (1) the plasma temperature reaches values below \SI{1}{\mega\kelvin}, where the response functions of the AIA EUV channels behave similarly; or (2) the temperatures do reach values above \SI{1}{\mega\kelvin}, but the short cooling timescales combined with the limitation of the AIA cadence (\SI{12}{} to \SI{6}{\second}) imply that the delays are not resolved. From our simulations, we suggest another interpretation (3) for the short delays: the impact of combined high density effects and plasma flows resulting from impulsive heating.

Interpretation (1) is the explanation for the short delays between the intensity peaks in m$_1$, m$_2$ (Fig. \ref{fig:results:lc:CL_HL_groupI_these}) and m$_{14}$ (Fig. \ref{fig:results:lc:CL_HL_groupeIII_these}). In fact, none of these three cool loops evolve into a hot loop (Fig. \ref{fig:results:m124_these_nt_evolution_v3}). Therefore, the density profile along the loop will remain the same, and the short delays can be explained by (1). On the other hand, interpretation (2) is not verified in our simulations. The models with initial hot loops show delays above \SI{30}{\second} between the AIA 193 and AIA 171 intensity peaks (Fig. \ref{fig:results:lc:HL_HL_groupI_these}). In that case, the radiative and conductive cooling of the loops' coronal part is not sufficiently fast to produce time delays below the AIA cadence of \SI{12}{\second}. Of course, this conclusion is subject to our choice of radiative loss function, which is further discussed in Appendix \ref{sec:annex:profile_cl}.

With regard to interpretation (3), high density effects and plasma flows can also induce co-temporal intensity peaks among all the HRIEUV, AIA, and SPICE light curves. We showed this for models with loops in an initial cool state evolving into a hot state (Section \ref{sec:results:flows:cl}), and in one example of a loop in an initial hot state (m$_{35}$, Section \ref{sec:results:flows:hl}). Taking m$_4$ as an example (initial cool state), the light curves peak co-temporally for all channels in less than \SI{10}{\second} after impulsive heating because of the rapid heating to $\log{T} \geq 5.0$ and of high density all along the loop. Then, the intensities of all channels decrease co-temporally for about \SI{50}{\second}, while the density of the loop decreases because of plasma flows (around \SI{20}{\kilo\meter\per\second}). This is the period when the plasma is ejected out of the loop into the chromosphere, and when the initial cool loop evolves into a final hot loop. We note that the timescale of this period (\SI{50}{\second}) is inferior to the radiative cooling timescale when the loop is heated up to $\log{T} = 5.8$ ($\approx$ \SI{500}{\second}), but is similar to the evacuation timescale of the loop ($\approx$ \SI{75}{\second} for a \SI{20}{\kilo\meter\per\second} flow). This highlights the importance of plasma flows and density effects for the light curve behavior in models with loops in an initial cool state and evolving into a hot state. As for m$_{35}$, the formation of downward propagating shock fronts and evaporating upflows (up to \SI{100}{\kilo\meter\per\second}) are responsible for the two co-temporal intensity peaks between some channels (Section \ref{sec:results:flows:hl}). Similar dense fronts and flows are also present for models with loops in an initial hot state other than m$_{35}$, but with lower front density and flow velocities. Their impact on the light curves behavior is therefore reduced, but not null. For example, in the case of m$_{22}$, they are responsible for the co-temporal peak seen in the \ion{C}{iii}, \ion{O}{iv}, and \ion{O}{vi} light curves at $t=$ \SI{108}{\second} (Fig. \ref{fig:results:lc:HL_HL_groupI_these}).  

In the observations, flows associated with events detected in HRIEUV have been previously reported. Using IRIS, \cite{Nelson_2023} measured $\approx$ \SI{22}{\kilo\meter\per\second} Doppler velocities with chromospheric lines ($\sim$ \SI{e4}{\kelvin}) on QS events. This is in line with the $\approx$ \SI{23}{\kilo\meter\per\second} flows in m$_{4}$ resulting from the evolution of a cool loop into a hot loop (Section \ref{sec:results:flows:cl}). Furthermore, the simulated flows are associated with plasma at temperatures ranging between $\log{T}$ = 4.5 and 5.5 (Fig. \ref{fig:results:CL_HL_m4_lc_ntv}), which is consistent with the emission temperature of the lines measured with IRIS. 

We can also compare our results with EEs in the QS \citep[]{Teriaca_2004}, investigated with the Solar Ultraviolet Measurements of Emitted Radiation \citep[SUMER,][]{Wilhelm1995} on board the Solar and Heliospheric Observatory \citep[SOHO,][]{Domingo_1995}. Explosive events are small ($\approx$ \SI{1.6}{\mega\meter}) and short-lived ($\approx$ \SI{60}{\second}) impulsive emissions detected in the \ion{O}{vi} \SI{1031.9}{\angstrom} line ($\log{T} = 5.6$), with Doppler velocities up to $\approx$ \SI{100}{\kilo\meter\per\second}. Their properties are very similar to those of the events detected by HRIEUV \citep[][]{Berghmans2021}. Thus, it is largely believed that part of our EUV events are EEs \citep[][]{Nelson_2023}. In our case, Doppler velocities measured in EEs are consistent with the flows of plasma at TR temperatures in m$_{35}$, up to \SI{100}{\kilo\meter\per\second}. Thus, we suggest that EEs may originate from short loops in a hot state subjected to a strong impulsive heating. This is in line with the other models, such as the one proposed by \cite{Peter_2019}: a small magnetic patch reconnecting with larger field lines of opposite polarity.  

\section{Conclusion}
\label{sec:conclusion}

In this work, we interpret the QS EUV brightenings detected in HRIEUV as the result of impulsive plasma heating in short loops. We provide models with loops in an initial cool ($T < \SI{1e5}{\kelvin}$) or hot state ($T > $ \SI{1e5}{\kelvin}). We find that models with impulsively heated cool loops reproduce well the light curve behavior of the events observed in HRIEUV, SPICE, and AIA. Thus, we suggest that they are good candidates for explaining their physical origin. On the other hand, impulsively heated hot loops models are not consistent with observations, except when the loop is subjected to strong impulsive heating (corresponding to an injected energy of \SI{2.4}{\erg\per\centi\meter\tothe{3}}). This model is also a promising lead for explaining the origin of some events, including those with two co-temporal intensity peaks and those within a short loop already visible in HRIEUV prior to the events themselves.

We also tested the cases of loops with an elliptical shape (group IV, Appendix \ref{annex:sec:groupIV_results}) and impulsive heatings with a narrow distribution function, located at the apex of the loop or at only one footpoint (group V, Appendix \ref{annex:sec:groupV_results}). The light curves resulting from group IV models do not show any significant differences compared to those of loops with a semicircular geometry. On the other hand, the results from group V (strong and narrow heating) show that the intensity peaks of the SPICE light curve can have time delays up to \SI{25}{\second}, if the initial cool loop is subjected to an impulsive heating with a narrow distribution function localized near the chromosphere. This suggests that the impulsive heating at the origin of EUV brightenings might be localized near the apex of the loop. However, the cadence of the SPICE light curves is also equal to \SI{25}{\second} in \cite{Dolliou_2024}. Such time delays between the intensity peaks might not be temporally resolved. As a reminder, this conclusion is based on the assumption that the intensities are averaged over the whole loop. Given a resolved loop and a heating localized on one of the footpoints, the light curve results could be very different, depending on where the LOS is.

We evaluated whether the behavior of the HRIEUV, AIA, and SPICE light curves is a good diagnostic to definitively test if the events reach coronal temperatures. We show that this is not the case (Fig.\ref{fig:results:fig1_tmin_tmax}). Indeed, some of the models with light curves showing co-temporal intensity peaks can originate from loops reaching coronal temperatures (m$_7$, m$_{16}$, m$_{35}$), while others originate from loops not reaching \SI{1}{\mega\kelvin} (\textit{e.g.}, m$_1$ to m$_4$). This result is also supported by the behavior of the hotter light curve (\ion{Mg}{ix}), which closely mimics the observations.

In conclusion, measuring the delays between the intensity peaks of the HRIEUV, SPICE, and AIA light curves is an effective diagnostic to determine the thermal model of the initial state of the loop (cool or hot). However, by itself, it is not a sufficient diagnostic to state whether the loop reaches or does not reach coronal temperatures \citep[see also][]{Chen_2025}. It should be combined with measurements of Doppler velocity to assess whether plasma flows play a role in the observed co-temporal intensity peaks. Additionally, temperature diagnostics with spectroscopy measuring a wide range of TR to coronal lines is preferred when possible. The \ion{Mg}{ix} line measured with SPICE is weak in the QS, so no emission can be detected even if the loop reaches temperatures above $\log{T} \geq 6.0$. As of now, only EIS can provide strong coronal lines in the QS. Therefore, it is the only instrument that can precisely measure the event emission at $\log{T} > 6.0$. However, the spatial resolution ($3^{\prime \prime}$) and exposure time of EIS \citep[\SI{60}{\second} in][]{Dolliou_2024} are strong limitations. Indeed, small events are generally not resolved by EIS and have a low filling factor. Therefore, they might remain undetected in the coronal lines of EIS, even if they do reach coronal temperatures. In the long term, these limitations should be addressed by the future Solar-C/EUV High-throughput Spectroscopic Telescope \citep[EUVST,][]{Shimizu_2019}, as it will reach higher spatial ($0.4^{\prime\prime}$) and temporal resolutions (\SI{1}{\second}). The lines measured by EUVST are also expected to cover plasma temperatures from the chromosphere ($\sim$ \SI{e4}{\kelvin}) to the corona ($\sim$ \SI{e6}{\kelvin}). Coordinated sequences including HRIEUV and EUVST will be well suited to measure the events' temperatures. Spectroscopic diagnostics applied to a large sample of events will allow us to evaluate the proportion of events that reach coronal temperatures with more precision. This is necessary to understand their contribution to the total observed coronal emission. Furthermore, their impact on TR heating must also be evaluated in future studies.
\begin{acknowledgements}
The authors thank the anonymous referee for the useful comments that helped to improve the manuscript. The authors gratefully thank S. Bradshaw, J. Reep and W. T. Barnes for the fruitful discussions on HYDRAD and forward modeling methods. A.D. acknowledges funding at IAS by CNES and EDOM (Université Paris-Saclay) through the Ph.D. Scholarships, and by CNES through the MEDOC data and operations center. At MPS, the work of A. Dolliou is funded by the Federal Ministry for Economic Affairs and Climate Action (BMWK) through the German Space Agency at DLR based on a decision of the German Bundestag (Funding code: 50OU2101, 50OU2201). J.A.K. was supported by the Heliophysics Internal Scientist Funding Model (HISFM) competitive grant program. This research was supported by the International Space Science Institute (ISSI) in Bern, through ISSI International Team project \#23-586 (Novel Insights Into Bursts, Bombs, and Brightenings in the Solar Atmosphere from Solar Orbiter). This research used CHIANTI version 10.1. CHIANTI is a collaborative project involving George Mason University, the University of Michigan (USA), University of Cambridge (UK) and NASA Goddard Space Flight Center (USA). This research used version 0.7.4 \citep{barnes_2021_5606094} of the aiapy open source software package \citep{Barnes2020}.

\end{acknowledgements}

\bibliographystyle{aa}
\bibliography{Biblio.bib}

\begin{thebibliography}{93}
\expandafter\ifx\csname natexlab\endcsname\relax\def\natexlab#1{#1}\fi

\bibitem[{{Anderson} {et~al.}(2020){Anderson}, {Appourchaux}, {Auchère},
  {Aznar Cuadrado}, {Barbay}, {Baudin}, {Beardsley}, {Bocchialini}, {Borgo}, \&
  et~al.}]{Anderson_2020}
{Anderson}, M., {Appourchaux}, T., {Auchère}, F., {et~al.} 2020, \aap, 642,
  A14

\bibitem[{{Antiochos} \& {Noci}(1986)}]{Antiochos_1986}
{Antiochos}, S.~K. \& {Noci}, G. 1986, \apj, 301, 440

\bibitem[{{Aschwanden} \& {Parnell}(2002)}]{Aschwanden_2002}
{Aschwanden}, M.~J. \& {Parnell}, C.~E. 2002, \apj, 572, 1048

\bibitem[{{Asplund} {et~al.}(2021){Asplund}, {Amarsi}, \&
  {Grevesse}}]{Asplund_2021}
{Asplund}, M., {Amarsi}, A.~M., \& {Grevesse}, N. 2021, \aap, 653, A141

\bibitem[{{Athay}(1986)}]{Athay_1986}
{Athay}, R.~G. 1986, \apj, 308, 975

\bibitem[{{Barczynski} {et~al.}(2017){Barczynski}, {Peter}, \&
  {Savage}}]{Barczynski_2017}
{Barczynski}, K., {Peter}, H., \& {Savage}, S.~L. 2017, \aap, 599, A137

\bibitem[{Barnes {et~al.}(2021)Barnes, Cheung, Bobra, Boerner, Chintzoglou,
  Freij, Leonard, Mumford, Padmanabhan, Shih, Shirman, Stansby, \&
  Wright}]{barnes_2021_5606094}
Barnes, W., Cheung, M., Bobra, M., {et~al.} 2021, aiapy

\bibitem[{{Barnes} {et~al.}(2016{\natexlab{a}}){Barnes}, {Cargill}, \&
  {Bradshaw}}]{Barnes_2016a}
{Barnes}, W.~T., {Cargill}, P.~J., \& {Bradshaw}, S.~J. 2016{\natexlab{a}},
  \apj, 829, 31

\bibitem[{{Barnes} {et~al.}(2016{\natexlab{b}}){Barnes}, {Cargill}, \&
  {Bradshaw}}]{Barnes_2016b}
{Barnes}, W.~T., {Cargill}, P.~J., \& {Bradshaw}, S.~J. 2016{\natexlab{b}},
  \apj, 833, 217

\bibitem[{Barnes {et~al.}(2020)Barnes, Cheung, Bobra, Boerner, Chintzoglou,
  Leonard, Mumford, Padmanabhan, Shih, Shirman, Stansby, \&
  Wright}]{Barnes2020}
Barnes, W.~T., Cheung, M. C.~M., Bobra, M.~G., {et~al.} 2020, Journal of Open
  Source Software, 5, 2801

\bibitem[{{Battaglia} {et~al.}(2021){Battaglia}, {Saqri}, {Massa},
  {Perracchione}, {Dickson}, {Xiao}, {Veronig}, {Warmuth}, {Battaglia},
  {Hurford}, {Meuris}, {Limousin}, {Etesi}, {Maloney}, {Schwartz}, {Kuhar},
  {Schuller}, {Senthamizh Pavai}, {Musset}, {Ryan}, {Kleint}, {Piana},
  {Massone}, {Benvenuto}, {Sylwester}, {Litwicka}, {St{\c{e}}{\'s}licki},
  {Mrozek}, {Vilmer}, {F{\'a}rn{\'\i}k}, {Ka{\v{s}}parov{\'a}}, {Mann},
  {Gallagher}, {Dennis}, {Csillaghy}, {Benz}, \& {Krucker}}]{Battaglia_2021}
{Battaglia}, A.~F., {Saqri}, J., {Massa}, P., {et~al.} 2021, \aap, 656, A4

\bibitem[{{Berghmans} {et~al.}(2021){Berghmans}, {Auch{\`e}re}, {Long},
  {Soubri{\'e}}, {Mierla}, {Zhukov}, {Sch{\"u}hle}, {Antolin}, {Harra},
  {Parenti}, {Podladchikova}, {Aznar Cuadrado}, {Buchlin}, {Dolla}, {Verbeeck},
  {Gissot}, {Teriaca}, {Haberreiter}, {Katsiyannis}, {Rodriguez}, {Kraaikamp},
  {Smith}, {Stegen}, {Rochus}, {Halain}, {Jacques}, {Thompson}, \&
  {Inhester}}]{Berghmans2021}
{Berghmans}, D., {Auch{\`e}re}, F., {Long}, D.~M., {et~al.} 2021, \aap, 656, L4

\bibitem[{{Berghmans} {et~al.}(1998{\natexlab{a}}){Berghmans}, {Clette}, \&
  {Moses}}]{Berghmans&Clette&Moses1998}
{Berghmans}, D., {Clette}, F., \& {Moses}, D. 1998{\natexlab{a}}, \aap, 336,
  1039

\bibitem[{{Berghmans} {et~al.}(1998{\natexlab{b}}){Berghmans}, {Clette}, \&
  {Moses}}]{Berghmans_1998}
{Berghmans}, D., {Clette}, F., \& {Moses}, D. 1998{\natexlab{b}}, \aap, 336,
  1039

\bibitem[{Bonet {et~al.}(2008)Bonet, Márquez, Almeida, Cabello, \&
  Domingo}]{Bonet_2008}
Bonet, J.~A., Márquez, I., Almeida, J.~S., Cabello, I., \& Domingo, V. 2008,
  The Astrophysical Journal, 687, L131

\bibitem[{{Bradshaw} \& {Klimchuk}(2011)}]{Bradshaw_Klimchuk_2011}
{Bradshaw}, S.~J. \& {Klimchuk}, J.~A. 2011, \apjs, 194, 26

\bibitem[{{Bradshaw} \& {Mason}(2003)}]{Bradshaw_2003}
{Bradshaw}, S.~J. \& {Mason}, H.~E. 2003, \aap, 401, 699

\bibitem[{{Cally} \& {Robb}(1991)}]{Cally_1991}
{Cally}, P.~S. \& {Robb}, T.~D. 1991, \apj, 372, 329

\bibitem[{{Cargill}(2014)}]{Cargill_2014}
{Cargill}, P.~J. 2014, \apj, 784, 49

\bibitem[{Cargill \& Klimchuk(1997)}]{Cargill_1997}
Cargill, P.~J. \& Klimchuk, J.~A. 1997, \apj, 478, 799

\bibitem[{{Chen} {et~al.}(2021){Chen}, {Przybylski}, {Peter}, {Tian},
  {Auch\`ere}, \& {Berghmans}}]{chen2021}
{Chen}, Y., {Przybylski}, D., {Peter}, H., {et~al.} 2021, A\&A, 656, L7

\bibitem[{{Chen, Yajie} {et~al.}(2025){Chen, Yajie}, {Peter, Hardi}, \&
  {Przybylski, Damien}}]{Chen_2025}
{Chen, Yajie}, {Peter, Hardi}, \& {Przybylski, Damien}. 2025, \aap, 693, A29

\bibitem[{{Cook} {et~al.}(1989){Cook}, {Cheng}, {Jacobs}, \&
  {Antiochos}}]{Cook_1989}
{Cook}, J.~W., {Cheng}, C.~C., {Jacobs}, V.~L., \& {Antiochos}, S.~K. 1989,
  \apj, 338, 1176

\bibitem[{{Crosby} {et~al.}(1993){Crosby}, {Aschwanden}, \&
  {Dennis}}]{Crosby_1993}
{Crosby}, N.~B., {Aschwanden}, M.~J., \& {Dennis}, B.~R. 1993, \solphys, 143,
  275

\bibitem[{{Culhane} {et~al.}(2007){Culhane}, {Harra}, {James}, {Al-Janabi},
  {Bradley}, {Chaudry}, {Rees}, {Tandy}, {Thomas}, {Whillock}, {Winter},
  {Doschek}, {Korendyke}, {Brown}, {Myers}, {Mariska}, {Seely}, {Lang}, {Kent},
  {Shaughnessy}, {Young}, {Simnett}, {Castelli}, {Mahmoud}, {Mapson-Menard},
  {Probyn}, {Thomas}, {Davila}, {Dere}, {Windt}, {Shea}, {Hagood}, {Moye},
  {Hara}, {Watanabe}, {Matsuzaki}, {Kosugi}, {Hansteen}, \&
  {Wikstol}}]{Culhane_2007}
{Culhane}, J.~L., {Harra}, L.~K., {James}, A.~M., {et~al.} 2007, \solphys, 243,
  19

\bibitem[{{De Pontieu} {et~al.}(2014){De Pontieu}, {Title}, {Lemen}, {Kushner},
  {Akin}, {Allard}, {Berger}, {Boerner}, {Cheung}, {Chou}, {Drake}, {Duncan},
  {Freeland}, {Heyman}, {Hoffman}, {Hurlburt}, {Lindgren}, {Mathur}, {Rehse},
  {Sabolish}, {Seguin}, {Schrijver}, {Tarbell}, {W{\"u}lser}, {Wolfson},
  {Yanari}, {Mudge}, {Nguyen-Phuc}, {Timmons}, {van Bezooijen}, {Weingrod},
  {Brookner}, {Butcher}, {Dougherty}, {Eder}, {Knagenhjelm}, {Larsen},
  {Mansir}, {Phan}, {Boyle}, {Cheimets}, {DeLuca}, {Golub}, {Gates}, {Hertz},
  {McKillop}, {Park}, {Perry}, {Podgorski}, {Reeves}, {Saar}, {Testa}, {Tian},
  {Weber}, {Dunn}, {Eccles}, {Jaeggli}, {Kankelborg}, {Mashburn}, {Pust},
  {Springer}, {Carvalho}, {Kleint}, {Marmie}, {Mazmanian}, {Pereira}, {Sawyer},
  {Strong}, {Worden}, {Carlsson}, {Hansteen}, {Leenaarts}, {Wiesmann},
  {Aloise}, {Chu}, {Bush}, {Scherrer}, {Brekke}, {Martinez-Sykora}, {Lites},
  {McIntosh}, {Uitenbroek}, {Okamoto}, {Gummin}, {Auker}, {Jerram}, {Pool}, \&
  {Waltham}}]{DePontieu2014}
{De Pontieu}, B., {Title}, A.~M., {Lemen}, J.~R., {et~al.} 2014, \solphys, 289,
  2733

\bibitem[{{Del Zanna} {et~al.}(2021){Del Zanna}, {Dere}, {Young}, \&
  {Landi}}]{Del_zanna_chianti_2021}
{Del Zanna}, G., {Dere}, K.~P., {Young}, P.~R., \& {Landi}, E. 2021, \apj, 909,
  38

\bibitem[{{Dere} {et~al.}(2023){Dere}, {Del Zanna}, {Young}, \&
  {Landi}}]{Dere_2023}
{Dere}, K.~P., {Del Zanna}, G., {Young}, P.~R., \& {Landi}, E. 2023, \apjs,
  268, 52

\bibitem[{{Dere} {et~al.}(1997){Dere}, {Landi}, {Mason}, {Monsignori Fossi}, \&
  {Young}}]{Dere_Chianti}
{Dere}, K.~P., {Landi}, E., {Mason}, H.~E., {Monsignori Fossi}, B.~C., \&
  {Young}, P.~R. 1997, \aaps, 125, 149

\bibitem[{{Dere} {et~al.}(2009){Dere}, {Landi}, {Young}, {Del Zanna},
  {Landini}, \& {Mason}}]{Dere_2009}
{Dere}, K.~P., {Landi}, E., {Young}, P.~R., {et~al.} 2009, \aap, 498, 915

\bibitem[{{Dolliou} {et~al.}(2023){Dolliou}, {Parenti, S.}, {Auchère, F.},
  {Bocchialini, K.}, {Pelouze, G.}, {Antolin, P.}, {Berghmans, D.}, {Harra,
  L.}, {Long, D. M.}, {Schühle, U.}, {Kraaikamp, E.}, {Stegen, K.}, {Verbeeck,
  C.}, {Gissot, S.}, {Aznar Cuadrado, R.}, {Buchlin, E.}, {Mierla, M.},
  {Teriaca, L.}, \& {Zhukov, A. N.}}]{Dolliou_2023}
{Dolliou}, A., {Parenti, S.}, {Auchère, F.}, {et~al.} 2023, \aa, 671, A64

\bibitem[{{Dolliou} {et~al.}(2024){Dolliou}, {Parenti, S.}, \& {Bocchialini,
  K.}}]{Dolliou_2024}
{Dolliou}, A., {Parenti, S.}, \& {Bocchialini, K.} 2024, \aa, 688, A77

\bibitem[{{Domingo} {et~al.}(1995){Domingo}, {Fleck}, \&
  {Poland}}]{Domingo_1995}
{Domingo}, V., {Fleck}, B., \& {Poland}, A.~I. 1995, \solphys, 162, 1

\bibitem[{Feldman {et~al.}(1999)Feldman, Widing, \& Warren}]{Feldman_1999}
Feldman, U., Widing, K.~G., \& Warren, H.~P. 1999, \apj, 522, 1133

\bibitem[{{Gudiksen} \& {Nordlund}(2002)}]{Gudiksen_2002}
{Gudiksen}, B.~V. \& {Nordlund}, {\r{A}}. 2002, \apjl, 572, L113

\bibitem[{{Gudiksen} \& {Nordlund}(2005{\natexlab{a}})}]{Gudiksen_2005a}
{Gudiksen}, B.~V. \& {Nordlund}, {\r{A}}. 2005{\natexlab{a}}, \apj, 618, 1031

\bibitem[{{Gudiksen} \& {Nordlund}(2005{\natexlab{b}})}]{Gudiksen_2005b}
{Gudiksen}, B.~V. \& {Nordlund}, {\r{A}}. 2005{\natexlab{b}}, \apj, 618, 1020

\bibitem[{{Hannah} {et~al.}(2008){Hannah}, {Christe}, {Krucker}, {Hurford},
  {Hudson}, \& {Lin}}]{Hannah_2008}
{Hannah}, I.~G., {Christe}, S., {Krucker}, S., {et~al.} 2008, \apj, 677, 704

\bibitem[{{Hannah} {et~al.}(2019){Hannah}, {Kleint}, {Krucker}, {Grefenstette},
  {Glesener}, {Hudson}, {White}, \& {Smith}}]{hannah_2019}
{Hannah}, I.~G., {Kleint}, L., {Krucker}, S., {et~al.} 2019, \apj, 881, 109

\bibitem[{{Hansteen} {et~al.}(2014){Hansteen}, {De Pontieu}, {Carlsson},
  {Lemen}, {Title}, {Boerner}, {Hurlburt}, {Tarbell}, {Wuelser}, {Pereira}, {De
  Luca}, {Golub}, {McKillop}, {Reeves}, {Saar}, {Testa}, {Tian}, {Kankelborg},
  {Jaeggli}, {Kleint}, \& {Mart{\'\i}nez-Sykora}}]{Hansteen2014}
{Hansteen}, V., {De Pontieu}, B., {Carlsson}, M., {et~al.} 2014, Science, 346,
  1255757

\bibitem[{{Huang} {et~al.}(2023){Huang}, {Teriaca.}, {Aznar Cuadrado, R.},
  {Chitta, L. P.}, {Mandal, S.}, {Peter, H.}, {Schühle, U.}, {Solanki, S. K.},
  {Auchère, F.}, {Berghmans, D.}, {Buchlin, É.}, {Carlsson, M.}, {Fludra,
  A.}, {Fredvik, T.}, {Giunta, A.}, {Grundy, T.}, {Hassler, D.}, {Parenti, S.},
  \& {Plaschke, F.}}]{Huang_2023}
{Huang}, Z., {Teriaca.}, L., {Aznar Cuadrado, R.}, {et~al.} 2023, \aa, 673, A82

\bibitem[{{Hudson}(1991)}]{Hudson1991}
{Hudson}, H.~S. 1991, \solphys, 133, 357

\bibitem[{{Joulin} {et~al.}(2016){Joulin}, {Buchlin}, {Solomon}, \&
  {Guennou}}]{Joulin_2016}
{Joulin}, V., {Buchlin}, E., {Solomon}, J., \& {Guennou}, C. 2016, \aap, 591,
  A148

\bibitem[{{Kahil} {et~al.}(2022){Kahil}, {Hirzberger}, {Solanki}, {Chitta},
  {Peter}, {Auch{\`e}re}, {Sinjan}, {Orozco Su{\'a}rez}, {Albert}, {Albelo
  Jorge}, {Appourchaux}, {Alvarez-Herrero}, {Blanco Rodr{\'\i}guez},
  {Gandorfer}, {Germerott}, {Guerrero}, {Guti{\'e}rrez M{\'a}rquez}, {Kolleck},
  {del Toro Iniesta}, {Volkmer}, {Woch}, {Fiethe}, {G{\'o}mez Cama},
  {P{\'e}rez-Grande}, {Sanchis Kilders}, {Balaguer Jim{\'e}nez}, {Bellot
  Rubio}, {Calchetti}, {Carmona}, {Deutsch}, {Fern{\'a}ndez-Rico},
  {Fern{\'a}ndez-Medina}, {Garc{\'\i}a Parejo}, {Gasent-Blesa}, {Gizon},
  {Grauf}, {Heerlein}, {Lagg}, {Lange}, {L{\'o}pez Jim{\'e}nez}, {Maue},
  {Meller}, {Michalik}, {Moreno Vacas}, {M{\"u}ller}, {Nakai}, {Schmidt},
  {Schou}, {Sch{\"u}hle}, {Staub}, {Strecker}, {Torralbo}, {Valori}, {Aznar
  Cuadrado}, {Teriaca}, {Berghmans}, {Verbeeck}, {Kraaikamp}, \&
  {Gissot}}]{Kahil_2022}
{Kahil}, F., {Hirzberger}, J., {Solanki}, S.~K., {et~al.} 2022, \aap, 660, A143

\bibitem[{{Klimchuk}(2006)}]{Klimchuk_2006}
{Klimchuk}, J.~A. 2006, \solphys, 234, 41

\bibitem[{Klimchuk(2015)}]{Klimchuk_2015}
Klimchuk, J.~A. 2015, Philosophical Transactions of the Royal Society A:
  Mathematical, Physical and Engineering Sciences, 373, 20140256

\bibitem[{{Klimchuk} {et~al.}(1987){Klimchuk}, {Antiochos}, \&
  {Mariska}}]{Klimchuk_1987}
{Klimchuk}, J.~A., {Antiochos}, S.~K., \& {Mariska}, J.~T. 1987, \apj, 320, 409

\bibitem[{{Klimchuk} {et~al.}(2010){Klimchuk}, {Karpen}, \&
  {Antiochos}}]{Klimchuk_2010}
{Klimchuk}, J.~A., {Karpen}, J.~T., \& {Antiochos}, S.~K. 2010, \apj, 714, 1239

\bibitem[{{Klimchuk} \& {Luna}(2019)}]{Klimchuk_Luna_2019}
{Klimchuk}, J.~A. \& {Luna}, M. 2019, \apj, 884, 68

\bibitem[{{Klimchuk} \& {Mariska}(1988)}]{Klimchuk_Mariska_1988}
{Klimchuk}, J.~A. \& {Mariska}, J.~T. 1988, \apj, 328, 334

\bibitem[{{Klimchuk} {et~al.}(2008){Klimchuk}, {Patsourakos}, \&
  {Cargill}}]{Klimchuk_2008}
{Klimchuk}, J.~A., {Patsourakos}, S., \& {Cargill}, P.~J. 2008, \apj, 682, 1351

\bibitem[{{Kosugi} {et~al.}(2007){Kosugi}, {Matsuzaki}, {Sakao}, {Shimizu},
  {Sone}, {Tachikawa}, {Hashimoto}, {Minesugi}, {Ohnishi}, {Yamada}, {Tsuneta},
  {Hara}, {Ichimoto}, {Suematsu}, {Shimojo}, {Watanabe}, {Shimada}, {Davis},
  {Hill}, {Owens}, {Title}, {Culhane}, {Harra}, {Doschek}, \&
  {Golub}}]{Kosugi_2007}
{Kosugi}, T., {Matsuzaki}, K., {Sakao}, T., {et~al.} 2007, \solphys, 243, 3

\bibitem[{{Kuniyoshi} {et~al.}(2024){Kuniyoshi}, {Bose}, \&
  {Yokoyama}}]{Kuniyoshi_2024}
{Kuniyoshi}, H., {Bose}, S., \& {Yokoyama}, T. 2024, \apjl, 969, L34

\bibitem[{{Lemen} {et~al.}(2012){Lemen}, {Title}, {Akin}, {Boerner}, {Chou},
  {Drake}, {Duncan}, {Edwards}, {Friedlaender}, {Heyman}, {Hurlburt}, {Katz},
  {Kushner}, {Levay}, {Lindgren}, {Mathur}, {McFeaters}, {Mitchell}, {Rehse},
  {Schrijver}, {Springer}, {Stern}, {Tarbell}, {Wuelser}, {Wolfson}, {Yanari},
  {Bookbinder}, {Cheimets}, {Caldwell}, {Deluca}, {Gates}, {Golub}, {Park},
  {Podgorski}, {Bush}, {Scherrer}, {Gummin}, {Smith}, {Auker}, {Jerram},
  {Pool}, {Soufli}, {Windt}, {Beardsley}, {Clapp}, {Lang}, \&
  {Waltham}}]{Lemen2012}
{Lemen}, J.~R., {Title}, A.~M., {Akin}, D.~J., {et~al.} 2012, \solphys, 275, 17

\bibitem[{{McClymont} \& {Canfield}(1983)}]{McClymont_1983}
{McClymont}, A.~N. \& {Canfield}, R.~C. 1983, \apj, 265, 497

\bibitem[{{Moriyasu} {et~al.}(2004){Moriyasu}, {Kudoh}, {Yokoyama}, \&
  {Shibata}}]{Moriyasu_2004}
{Moriyasu}, S., {Kudoh}, T., {Yokoyama}, T., \& {Shibata}, K. 2004, \apjl, 601,
  L107

\bibitem[{{M\"uller} {et~al.}(2020){M\"uller}, {St. Cyr, O. C.}, {Zouganelis,
  I.}, {Gilbert, H. R.}, {Marsden, R.}, {Nieves-Chinchilla, T.}, {Antonucci,
  E.}, {Auch\`ere, F.}, {Berghmans, D.}, {Horbury, T. S.}, {Howard, R. A.},
  {Krucker, S.}, {Maksimovic, M.}, {Owen, C. J.}, {Rochus, P.},
  {Rodriguez-Pacheco, J.}, {Romoli, M.}, {Solanki, S. K.}, {Bruno, R.},
  {Carlsson, M.}, {Fludra, A.}, {Harra, L.}, {Hassler, D. M.}, {Livi, S.},
  {Louarn, P.}, {Peter, H.}, {Sch\"uhle, U.}, {Teriaca, L.}, {del Toro Iniesta,
  J. C.}, {Wimmer-Schweingruber, R. F.}, {Marsch, E.}, {Velli, M.}, {De Groof,
  A.}, {Walsh, A.}, \& {Williams, D.}}]{Muller}
{M\"uller}, D., {St. Cyr, O. C.}, {Zouganelis, I.}, {et~al.} 2020, A\&A, 642,
  A1

\bibitem[{{Nelson} {et~al.}(2023){Nelson}, {Auch{\`e}re}, {Aznar Cuadrado},
  {Barczynski}, {Buchlin}, {Harra}, {Long}, {Parenti}, {Peter}, {Sch{\"u}hle},
  {Schwanitz}, {Smith}, {Teriaca}, {Verbeeck}, {Zhukov}, \&
  {Berghmans}}]{Nelson_2023}
{Nelson}, C.~J., {Auch{\`e}re}, F., {Aznar Cuadrado}, R., {et~al.} 2023, \aap,
  676, A64

\bibitem[{{Nelson, C. J.} {et~al.}(2024){Nelson, C. J.}, {Hayes, L. A.},
  {Müller, D.}, {Musset, S.}, {Freij, N.}, {Auchère, F.}, {Aznar Cuadrado,
  R.}, {Barczynski, K.}, {Buchlin, E.}, {Harra, L.}, {Long, D. M.}, {Parenti,
  S.}, {Peter, H.}, {Schühle, U.}, {Smith, P.}, {Teriaca, L.}, {Verbeeck, C.},
  {Zhukov, A. N.}, \& {Berghmans, D.}}]{Nelson_2024}
{Nelson, C. J.}, {Hayes, L. A.}, {Müller, D.}, {et~al.} 2024, \aa, 692, A236

\bibitem[{Panesar {et~al.}(2021)Panesar, Tiwari, Berghmans, Cheung, Müller,
  Auchere, \& Zhukov}]{Panesar_2021}
Panesar, N.~K., Tiwari, S.~K., Berghmans, D., {et~al.} 2021, \apjl, 921, L20

\bibitem[{{Parenti} \& {Vial}(2007)}]{Parenti_2007}
{Parenti}, S. \& {Vial}, J.~C. 2007, \aap, 469, 1109

\bibitem[{{Parker}(1988)}]{Parker1988}
{Parker}, E.~N. 1988, \apj, 330, 474

\bibitem[{{Pesnell} {et~al.}(2012){Pesnell}, {Thompson}, \&
  {Chamberlin}}]{Pesnell2012}
{Pesnell}, W.~D., {Thompson}, B.~J., \& {Chamberlin}, P.~C. 2012, \solphys,
  275, 3

\bibitem[{{Peter}(2001)}]{Peter_2001}
{Peter}, H. 2001, \aap, 374, 1108

\bibitem[{{Peter} {et~al.}(2004){Peter}, {Gudiksen}, \&
  {Nordlund}}]{Peter_2004}
{Peter}, H., {Gudiksen}, B.~V., \& {Nordlund}, {\r{A}}. 2004, \apjl, 617, L85

\bibitem[{{Peter} {et~al.}(2019){Peter}, {Huang}, {Chitta}, \&
  {Young}}]{Peter_2019}
{Peter}, H., {Huang}, Y.~M., {Chitta}, L.~P., \& {Young}, P.~R. 2019, \aap,
  628, A8

\bibitem[{{Qiu} {et~al.}(2013){Qiu}, {Sturrock}, {Longcope}, {Klimchuk}, \&
  {Liu}}]{Qiu_2013}
{Qiu}, J., {Sturrock}, Z., {Longcope}, D.~W., {Klimchuk}, J.~A., \& {Liu},
  W.-J. 2013, \apj, 774, 14

\bibitem[{{Reale}(2014)}]{Reale2014}
{Reale}, F. 2014, Living Reviews in Solar Physics, 11, 4

\bibitem[{{Rempel}(2017)}]{Rempel_2017}
{Rempel}, M. 2017, \apj, 834, 10

\bibitem[{{Rochus} {et~al.}(2020)}]{EUI_instrument}
{Rochus}, P. {et~al.} 2020, \aap, 642, A8

\bibitem[{Rodríguez-Gómez {et~al.}(2024)Rodríguez-Gómez, Kuckein,
  González~Manrique, Saqri, Veronig, Gömöry, \&
  Podladchikova}]{Rodriguez_Gomez_2024}
Rodríguez-Gómez, J.~M., Kuckein, C., González~Manrique, S.~J., {et~al.}
  2024, The Astrophysical Journal, 964, 27

\bibitem[{{Rosner} {et~al.}(1978){Rosner}, {Tucker}, \& {Vaiana}}]{Rosner_1978}
{Rosner}, R., {Tucker}, W.~H., \& {Vaiana}, G.~S. 1978, \apj, 220, 643

\bibitem[{{Sasso} {et~al.}(2012){Sasso}, {Andretta}, {Spadaro}, \&
  {Susino}}]{Sasso_2012}
{Sasso}, C., {Andretta}, V., {Spadaro}, D., \& {Susino}, R. 2012, \aap, 537,
  A150

\bibitem[{{Shimizu, T.} {et~al.}(2019){Shimizu, T.}, {Imada, S.}, {Kawate, T.},
  {Ichimoto, K.}, {Suematsu, Y.}, {Hara, H.}, {Katsukawa, Y.}, {Kubo, M.},
  {Toriumi, S.}, {Watanabe, T.}, {Yokoyama, T.}, {Korendyke, C. M.}, {Warren,
  H. P.}, {Tarbell, T.}, {De Pontieu, B.}, {Teriaca, L.}, {Sch{\"u}hle, U. H.},
  {Solanki, S.}, {Harra, L. K.}, {Matthews, S.}, {Fludra, A.}, {Auch{\`e}re,
  F.}, {Andretta, V.}, {Naletto, G.}, \& {Zhukov, A.}}]{Shimizu_2019}
{Shimizu, T.}, {Imada, S.}, {Kawate, T.}, {et~al.} 2019, in UV, X-Ray, and
  Gamma-Ray Space Instrumentation for Astronomy XXI, ed. O.~H. Siegmund, Vol.
  11118, International Society for Optics and Photonics (SPIE), 1111807

\bibitem[{{Skan} {et~al.}(2023){Skan}, {Danilovic}, {Leenaarts}, {Calvo}, \&
  {Rempel}}]{Skan_2023}
{Skan}, M., {Danilovic}, S., {Leenaarts}, J., {Calvo}, F., \& {Rempel}, M.
  2023, \aap, 672, A47

\bibitem[{{Sukarmadji} \& {Antolin}(2024)}]{Sukarmadji_2024}
{Sukarmadji}, A. R.~C. \& {Antolin}, P. 2024, \apjl, 961, L17

\bibitem[{Team {et~al.}(2019)Team, Al-Janabi, Antolin, Baker, Bellot~Rubio,
  Bradley, Brooks, Centeno, Culhane, Del~Zanna, Doschek, Fletcher, Hara, Harra,
  Hillier, Imada, Klimchuk, Mariska, Pereira, Reeves, Sakao, Sakurai, Shimizu,
  Shimojo, Shiota, Solanki, Sterling, Su, Suematsu, Tarbell, Tiwari, Toriumi,
  Ugarte-Urra, Warren, Watanabe, \& Young}]{Hinode_2019}
Team, H.~R., Al-Janabi, K., Antolin, P., {et~al.} 2019, Publications of the
  Astronomical Society of Japan, 71, R1

\bibitem[{{Teriaca} {et~al.}(2004){Teriaca}, {Banerjee}, {Falchi}, {Doyle}, \&
  {Madjarska}}]{Teriaca_2004}
{Teriaca}, L., {Banerjee}, D., {Falchi}, A., {Doyle}, J.~G., \& {Madjarska},
  M.~S. 2004, \aap, 427, 1065

\bibitem[{{Tilipman} {et~al.}(2023){Tilipman}, {Kazachenko}, {Tremblay},
  {Mili{\'c}}, {Mart{\'\i}nez Pillet}, \& {Rempel}}]{Tilipman_2023}
{Tilipman}, D., {Kazachenko}, M., {Tremblay}, B., {et~al.} 2023, \apj, 956, 83

\bibitem[{{Tiwari} {et~al.}(2022){Tiwari}, {Hansteen}, {De Pontieu}, {Panesar},
  \& {Berghmans}}]{Tiwari2022}
{Tiwari}, S.~K., {Hansteen}, V.~H., {De Pontieu}, B., {Panesar}, N.~K., \&
  {Berghmans}, D. 2022, \apj, 929, 103

\bibitem[{{Ugarte-Urra} {et~al.}(2019){Ugarte-Urra}, {Crump}, {Warren}, \&
  {Wiegelmann}}]{Ugarte_2019}
{Ugarte-Urra}, I., {Crump}, N.~A., {Warren}, H.~P., \& {Wiegelmann}, T. 2019,
  \apj, 877, 129

\bibitem[{{Upendran} \& {Tripathi}(2021)}]{Upendran2021}
{Upendran}, V. \& {Tripathi}, D. 2021, \apj, 916, 59

\bibitem[{{Van Doorsselaere} {et~al.}(2020){Van Doorsselaere}, {Srivastava},
  {Antolin}, {Magyar}, {Vasheghani Farahani}, {Tian}, {Kolotkov}, {Ofman},
  {Guo}, {Arregui}, {De Moortel}, \& {Pascoe}}]{VanDoorsselaere_2020}
{Van Doorsselaere}, T., {Srivastava}, A.~K., {Antolin}, P., {et~al.} 2020,
  \ssr, 216, 140

\bibitem[{{Viall} {et~al.}(2021){Viall}, {De Moortel}, {Downs}, {Klimchuk},
  {Parenti}, \& {Reale}}]{Viall21}
{Viall}, N.~M., {De Moortel}, I., {Downs}, C., {et~al.} 2021, in Solar Physics
  and Solar Wind, ed. N.~E. {Raouafi} \& A.~{Vourlidas}, Vol.~1, 35

\bibitem[{Viall \& Klimchuk(2015)}]{Viall_2015}
Viall, N.~M. \& Klimchuk, J.~A. 2015, \apj, 799, 58

\bibitem[{Viall \& Klimchuk(2017)}]{viall_survey_2017}
Viall, N.~M. \& Klimchuk, J.~A. 2017, \apj, 842, 108, aDS Bibcode:
  2017ApJ...842..108V

\bibitem[{{V{\"o}gler} {et~al.}(2005){V{\"o}gler}, {Shelyag}, {Sch{\"u}ssler},
  {Cattaneo}, {Emonet}, \& {Linde}}]{Vogler_2005}
{V{\"o}gler}, A., {Shelyag}, S., {Sch{\"u}ssler}, M., {et~al.} 2005, \aap, 429,
  335

\bibitem[{{Wedemeyer-B{\"o}hm} {et~al.}(2012){Wedemeyer-B{\"o}hm}, {Scullion},
  {Steiner}, {Rouppe van der Voort}, {de La Cruz Rodriguez}, {Fedun}, \&
  {Erd{\'e}lyi}}]{Wedemeyer-Bohm_2012}
{Wedemeyer-B{\"o}hm}, S., {Scullion}, E., {Steiner}, O., {et~al.} 2012, \nat,
  486, 505

\bibitem[{{Wilhelm} {et~al.}(1995){Wilhelm}, {Curdt}, {Marsch}, {Sch{\"u}hle},
  {Lemaire}, {Gabriel}, {Vial}, {Grewing}, {Huber}, {Jordan}, {Poland},
  {Thomas}, {K{\"u}hne}, {Timothy}, {Hassler}, \& {Siegmund}}]{Wilhelm1995}
{Wilhelm}, K., {Curdt}, W., {Marsch}, E., {et~al.} 1995, \solphys, 162, 189

\bibitem[{{Winebarger} {et~al.}(2013){Winebarger}, {Walsh}, {Moore}, {De
  Pontieu}, {Hansteen}, {Cirtain}, {Golub}, {Kobayashi}, {Korreck}, {DeForest},
  {Weber}, {Title}, \& {Kuzin}}]{Winebarger_2013}
{Winebarger}, A.~R., {Walsh}, R.~W., {Moore}, R., {et~al.} 2013, \apj, 771, 21

\bibitem[{Zhang \& Liu(2011)}]{Zhang_2011}
Zhang, J. \& Liu, Y. 2011, The Astrophysical Journal Letters, 741, L7

\bibitem[{{Zhukov} {et~al.}(2021){Zhukov}, {Mierla}, {Auch{\`e}re}, {Gissot},
  {Rodriguez}, {Soubri{\'e}}, {Thompson}, {Inhester}, {Nicula}, {Antolin},
  {Parenti}, {Buchlin}, {Barczynski}, {Verbeeck}, {Kraaikamp}, {Smith},
  {Stegen}, {Dolla}, {Harra}, {Long}, {Sch{\"u}hle}, {Podladchikova}, {Aznar
  Cuadrado}, {Teriaca}, {Haberreiter}, {Katsiyannis}, {Rochus}, {Halain},
  {Jacques}, \& {Berghmans}}]{Zhukov2021}
{Zhukov}, A.~N., {Mierla}, M., {Auch{\`e}re}, F., {et~al.} 2021, \aap, 656, A35

\bibitem[{{Zouganelis} {et~al.}(2020){Zouganelis}, {De Groof}, {Walsh},
  {Williams}, {M{\"u}ller}, {St Cyr}, {Auch{\`e}re}, {Berghmans}, {Fludra},
  {Horbury}, {Howard}, {Krucker}, {Maksimovic}, {Owen},
  {Rodr{\'\i}guez-Pacheco}, {Romoli}, {Solanki}, {Watson}, {Sanchez}, {Lefort},
  {Osuna}, {Gilbert}, {Nieves-Chinchilla}, {Abbo}, {Alexandrova},
  {Anastasiadis}, {Andretta}, {Antonucci}, {Appourchaux}, {Aran}, {Arge},
  {Aulanier}, {Baker}, {Bale}, {Battaglia}, {Bellot Rubio}, {Bemporad},
  {Berthomier}, {Bocchialini}, {Bonnin}, {Brun}, {Bruno}, {Buchlin},
  {B{\"u}chner}, {Bucik}, {Carcaboso}, {Carr}, {Carrasco-Bl{\'a}zquez},
  {Cecconi}, {Cernuda Cangas}, {Chen}, {Chitta}, {Chust}, {Dalmasse},
  {D'Amicis}, {Da Deppo}, {De Marco}, {Dolei}, {Dolla}, {Dudok de Wit}, {van
  Driel-Gesztelyi}, {Eastwood}, {Espinosa Lara}, {Etesi}, {Fedorov},
  {F{\'e}lix-Redondo}, {Fineschi}, {Fleck}, {Fontaine}, {Fox}, {Gandorfer},
  {G{\'e}not}, {Georgoulis}, {Gissot}, {Giunta}, {Gizon}, {G{\'o}mez-Herrero},
  {Gontikakis}, {Graham}, {Green}, {Grundy}, {Haberreiter}, {Harra}, {Hassler},
  {Hirzberger}, {Ho}, {Hurford}, {Innes}, {Issautier}, {James}, {Janitzek},
  {Janvier}, {Jeffrey}, {Jenkins}, {Khotyaintsev}, {Klein}, {Kontar},
  {Kontogiannis}, {Krafft}, {Krasnoselskikh}, {Kretzschmar}, {Labrosse},
  {Lagg}, {Landini}, {Lavraud}, {Leon}, {Lepri}, {Lewis}, {Liewer}, {Linker},
  {Livi}, {Long}, {Louarn}, {Malandraki}, {Maloney}, {Martinez-Pillet},
  {Martinovic}, {Masson}, {Matthews}, {Matteini}, {Meyer-Vernet}, {Moraitis},
  {Morton}, {Musset}, {Nicolaou}, {Nindos}, {O'Brien}, {Orozco Suarez},
  {Owens}, {Pancrazzi}, {Papaioannou}, {Parenti}, {Pariat}, {Patsourakos},
  {Perrone}, {Peter}, {Pinto}, {Plainaki}, {Plettemeier}, {Plunkett}, {Raines},
  {Raouafi}, {Reid}, {Retino}, {Rezeau}, {Rochus}, {Rodriguez},
  {Rodriguez-Garcia}, {Roth}, {Rouillard}, {Sahraoui}, {Sasso}, {Schou},
  {Sch{\"u}hle}, {Sorriso-Valvo}, {Soucek}, {Spadaro}, {Stangalini}, {Stansby},
  {Steller}, {Strugarek}, {{\v{S}}tver{\'a}k}, {Susino}, {Telloni}, {Terasa},
  {Teriaca}, {Toledo-Redondo}, {del Toro Iniesta}, {Tsiropoula}, {Tsounis},
  {Tziotziou}, {Valentini}, {Vaivads}, {Vecchio}, {Velli}, {Verbeeck},
  {Verdini}, {Verscharen}, {Vilmer}, {Vourlidas}, {Wicks},
  {Wimmer-Schweingruber}, {Wiegelmann}, {Young}, \& {Zhukov}}]{Zouganelis2020}
{Zouganelis}, I., {De Groof}, A., {Walsh}, A.~P., {et~al.} 2020, \aap, 642, A3

\end{thebibliography}

\begin{appendix}

\section{Results for groups II, IV, and V}
\label{annex:sec:results_groups}

In this section we present the results for the groups II (Table \ref{table:code:groupII}), IV (Table \ref{table:code:groupIV}) and V (Table \ref{table:code:groupV}) models. In Section \ref{sec:annex:tmax_nmax}, we show the temperature and the density reached by each model in the apex at the time of the temperature peak. In Section \ref{sec:annex:lc}, we comment the light curves of HRIEUV, four EUV channels of AIA, and five emission lines computed from the simulations outputs. 

\subsection{Electron temperature and density reached after impulsive heating}
\label{sec:annex:tmax_nmax}

\begin{figure*}
    \includegraphics{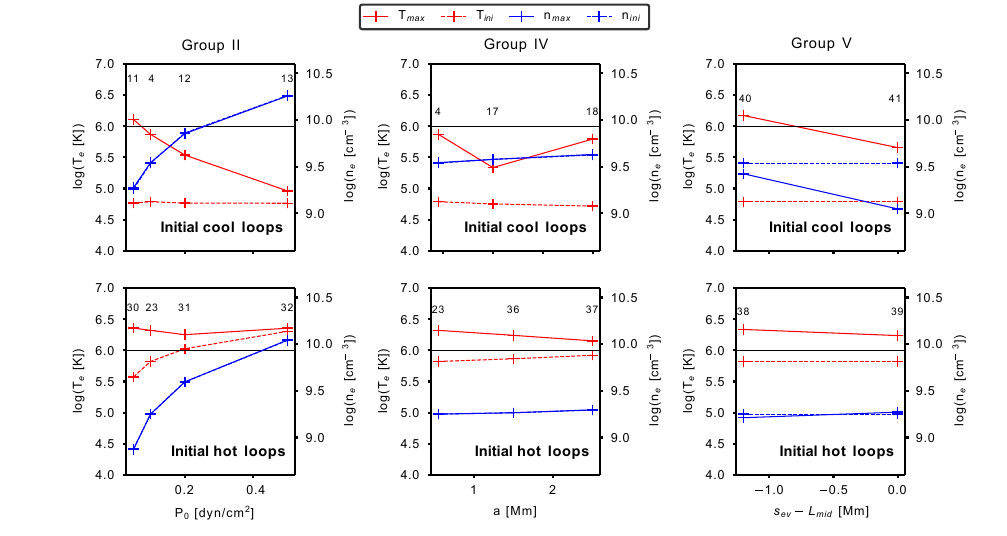}
    \caption{Maximal electron temperature  $T_\mathrm{max}$  and minimal electron density $n_\mathrm{min}$ reached during the simulation over the entire loop profile. Prior to the impulsive heating, the initial values $T_\mathrm{ini}$ and $n_\mathrm{ini}$ are also given when the loops were in equilibrium. Results for models from the groups II, IV and V are separated by columns from left to right, with the X-axis being the changing parameter associated with each group. Every models number given in tables \ref{table:code:groupI} to \ref{table:code:groupIII} (\textit{i.e.}, m$_1$ to m$_{35}$) are displayed on top of their respective data points (crosses), either for an initial cool (top row) or hot loop (bottom row).  The horizontal black line delimits the \SI{1}{\mega\kelvin} temperature. }
    \label{fig:annex:fig1_tmin_tmax}
\end{figure*}

For the groups II, IV and V models, Figure \ref{fig:annex:fig1_tmin_tmax} displays the electron temperature and density at the apex when the loop was initially at equilibrium ($T_\mathrm{ini}$, $n_\mathrm{ini}$), and at the times of the temperature peak after impulsive heating ($T_\mathrm{max}$, $n_\mathrm{max}$). We notice that all models with loops in an initial hot state have $T_\mathrm{max}$ above \SI{1}{\mega\kelvin}. Therefore, the impulsive heating associated with these models is likely to contribute to coronal emission above \SI{1}{\mega\kelvin}. As for models with loops in an initial cool state, most of the loops do not reach coronal temperatures. The exceptions are for loops with lower pressure (m$_{11}$, $P_0 = $ \SI{0.05}{\dyne\per\centi\meter\tothe{2}}) and for the loop in the m$_{40}$ model. In the latter, the impulsive heating has a higher maximal amplitude ($A_\mathrm{max} = $ \SI{0.79}{\erg\per\centi\meter\tothe{3}\per\second}) and is more narrowly distributed ($\sigma_\mathrm{ev} = 0.1L$ ) around the loop center than most other models ($A_\mathrm{max} = $ \SI{0.2}{\erg\per\centi\meter\tothe{3}\per\second} and $\sigma_\mathrm{ev} = 2L$ ). This is why it was to be expected that the loop in m$_{30}$ reaches higher temperatures at the apex ($\log{T_\mathrm{max}} = 6.2$) than other models with loops in an initial cool state.

\subsection{Light curves}
\label{sec:annex:lc}

In this section, we show the results of the light curves for models of group II (Section \ref{annex:sec:groupII_results}), group IV (Section \ref{annex:sec:groupIV_results}) and group V (Section \ref{annex:sec:groupV_results}). 

\subsubsection{Group II: Increasing the pressure $P_0$ at the top of the chromospheric legs}
\label{annex:sec:groupII_results}

\renewcommand{\arraystretch}{1.3}
\begin{table*}
\caption{Parameters of the group II models.}
\flushleft          
\begin{tabular}{c c c c | c c c | c | c c c c}   
\multicolumn{12}{c}{Group II} \\
\hline       
\hline       
\multicolumn{4}{c|}{Initial cool loop} & \multicolumn{3}{c|}{Initial hot loop} &
\multicolumn{1}{c|}{} & \multicolumn{4}{c}{Heating} \\
\hline       
  Name & $T_\mathrm{iCL}$ & $n_\mathrm{iCL}$ &  $P_0$  & Name & $T_\mathrm{iHL}$ & $n_\mathrm{iHL}$ & $L$ &  $H_0$ & $A_{\mathrm{max}}$ & $\sigma_\mathrm{ev}$ & iCL $\rightarrow$ fHL
\\
\hline
$m_{11}$ & 4.8 & 9.3 & 0.05 & $m_{30}$ & 5.6 & 8.9 & 3.0 & \SI{0.6e-3}{} & 0.2 & 6.0 & Yes  \\
$m_{4}$ & 4.8 & 9.5 & 0.1 & $m_{23}$ & 5.8 & 9.3 & & \SI{2.3e-3}{} & &  & Yes  \\
$m_{12}$ & 4.8 & 9.9 & 0.2 & $m_{31}$ & 6.0 & 9.6 & & \SI{9.3e-3}{} & & & Yes  \\
$m_{13}$ & 4.8 & 10.3 & 0.5 & $m_{32}$ & 6.3 & 10.0 & & \SI{58.2e-3}{} &  & & Yes  \\

\end{tabular}
\tablefoot{Initial parameters of the group II models, with variation of the pressure at the top of the vertical legs $P_{\mathrm{0}}$ for cool loops. As a consequence, the uniform and constant term $H_0$ of the heating also varies. The parameters are the same as for Table  \ref{table:code:groupI}.   } 
\label{table:code:groupII}      
\end{table*}

\begin{figure*}
    \centering
    \includegraphics{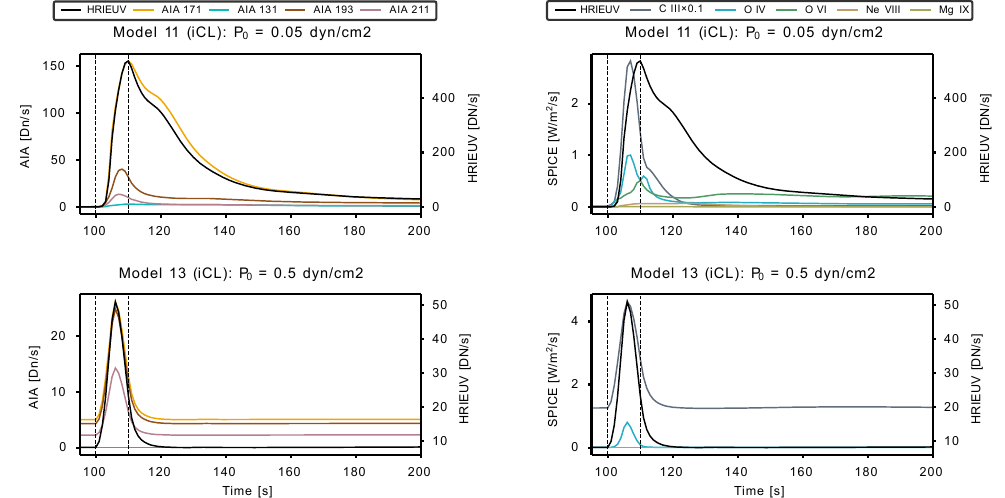}
    \includegraphics{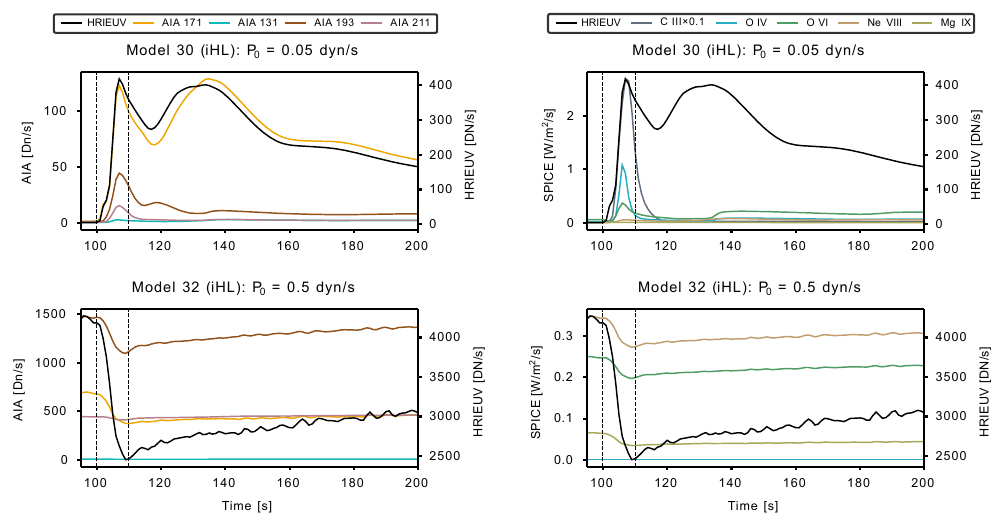}
    \caption{Light curves for the simulations of group II models (Table \ref{table:code:groupV}), with loops in an initial cool or hot states. The figure is similar to Fig.\ref{fig:results:lc:CL_HL_groupI_these}}
    \label{fig:annex:group2_light_curves}
\end{figure*}

Figure \ref{fig:annex:group2_light_curves} shows the light curves of group II models (Table \ref{table:code:groupII}), with an increasing pressure at the top of the vertical legs from $P_0 = $ \SI{0.05}{\dyne\per\centi\meter\tothe{2}} to \SI{0.5}{\dyne\per\centi\meter\tothe{2}}. We present the results for models with loops in an initial cool state (m$_{4}$ and m$_{18}$) and in an initial hot state (m$_{23}$ and m$_{37}$). Regarding m$_{4}$ and m$_{18}$, the results are very similar to those from  group I models with loops in an initial cool state (Fig. \ref{fig:results:lc:CL_HL_groupI_these}).

The model $m_{30}$ with a loop in an initial hot state and a low pressure ($P_0 =$ \SI{0.05}{\dyne\per\centi\meter\tothe{2}}) shows a co-temporal intensity peak at $t=$ \SI{107}{\second} between all light curves. We also notice another intensity peak by HRIEUV and AIA 171 around $t = $ \SI{135}{\second}. These intensity peaks are caused by flows similar to those in m$_{35}$ (Section \ref{sec:results:flows:hl}). Instead of resulting from a large amplitude in the impulsive heating (e.g.,m$_{35}$), the flows in m$_{30}$ are boosted by the lower pressure and viscosity. Therefore, increasing the impulsive heating amplitude or decreasing the pressure of a loop in a hot state has a similar impact to the light curves: both results in co-temporal intensity peaks. Finally, the HRIEUV, AIA 171, 193 and SPICE light curves of the model m$_{32}$ decrease co-temporally after impulsive heating. This is because the loop temperature slightly increases above $\log{T} = 6.2$ (Fig. \ref{fig:annex:fig1_tmin_tmax}), which is above the temperature peak of the response and contribution functions of most channels and lines (Fig. \ref{fig:annex:gofnt}). After $t=$ \SI{110}{\second}, the light curves slowly increases while the loop cools down through conduction and radiations. 

\subsubsection{Group IV: Increasing the elongation of semielliptical loops}
\label{annex:sec:groupIV_results}

\renewcommand{\arraystretch}{1.3}
\begin{table*}
\caption{Parameters of the group IV models.}
\flushleft          
\begin{tabular}{c c c c | c c c | c c c | c c c c}   
\multicolumn{14}{c}{Group IV} \\
\hline       
\hline       
\multicolumn{4}{c|}{Initial cool loop} & \multicolumn{3}{c|}{Initial hot loop} &
\multicolumn{3}{c|}{} & \multicolumn{4}{c}{Heating} \\
\hline       
  Name & $T_\mathrm{iCL}$ & $n_\mathrm{iCL}$ &  $P_0$  & Name & $T_\mathrm{iHL}$ & $n_\mathrm{iHL}$ & $L$ & $a$ & $b$&  $H_0$ & $A_{\mathrm{max}}$ & $\sigma_\mathrm{ev}$ & iCL $\rightarrow$ fHL
\\
\hline
$m_{4}$ & 4.8 & 9.5 & 0.1 & $m_{23}$ & 5.8 & 9.3 & 3.0 & 0.96 & 0.96 & 0.0023 & 0.2 & 6.0 & Yes  \\
$m_{17}$ & 4.8 & 9.6 & 0.1 & $m_{36}$ & 5.9 & 9.3 & 3.9 & 1.5 & &  &  & 7.9 & Yes  \\
$m_{18}$ & 4.8 & 9.6 & 0.1 & $m_{37}$ & 5.9 & 9.3 & 5.8 & 2.5 & &  &  & 11.7 & Yes  \\

\end{tabular}
\tablefoot{Initial parameters of the group IV models, with semielliptical central parts of semimajor axis $a$ and semiminor axis $b$ (\SI{}{\mega\meter}) along $z$ and $x$, respectively. The elongation $a$ varies, while $b = $ \SI{0.96}{\mega\meter} is constant for all models. The other parameters are the same as for Table  \ref{table:code:groupI}. } 
\label{table:code:groupIV}      
\end{table*}

\begin{figure*}
    \centering
    \includegraphics{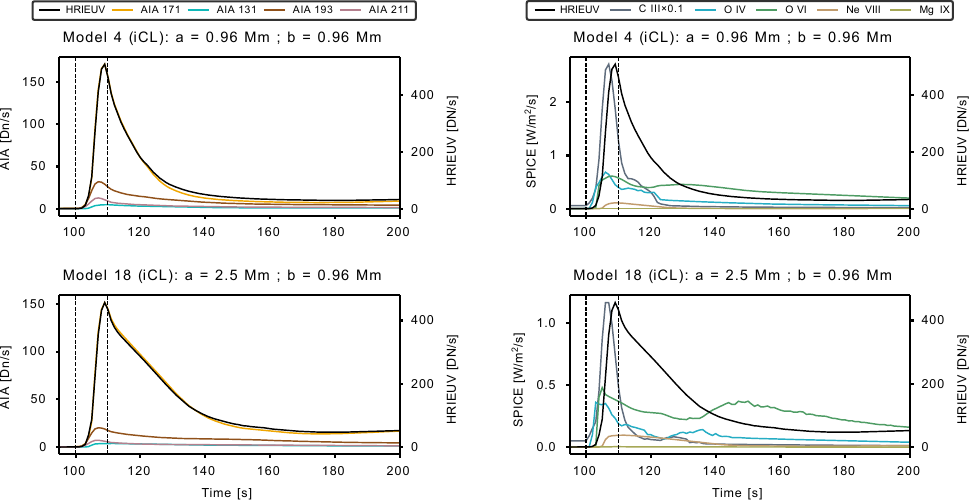}
    \includegraphics{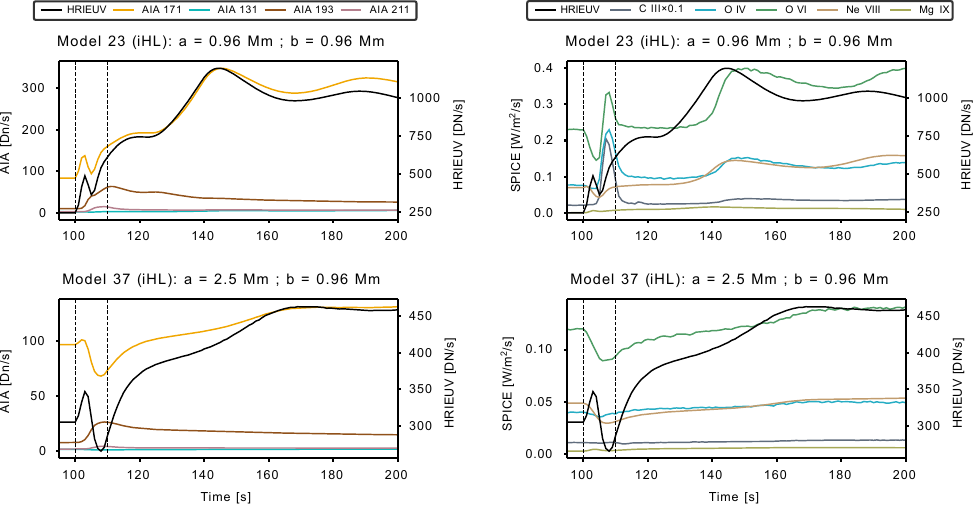}
    \caption{Light curves for the simulations of group IV models (Table \ref{table:code:groupV}), with loops in an initial cool or  hot state. The figure is similar to Fig.\ref{fig:results:lc:CL_HL_groupI_these}}
    \label{fig:annex:group4_light_curves}
\end{figure*}
We now present the results for the group IV models (Table \ref{table:code:groupIV}). Loops from these models have a semielliptical shape with two semiaxis $a$ and $b$ along $x$ and $z$, respectively (Section \ref{sec:code:loop_geometry}). In group IV, the elongation $a$ of the loop increases from $a=$ \SI{96}{\mega\meter} (m$_4$ and m$_{23}$) and to \SI{2.5}{\mega\meter} (m$_{18}$ and m$_{37}$), while the loop's height $b$ remains constant. The m$_{23}$ models corresponds to the semicircular case with $L = $ \SI{3}{\mega\meter}. 

Figure \ref{fig:annex:group4_light_curves} shows the light curves of group IV models with loops in an initial cool state and in an initial hot state. The results are similar to those from group I models (Section \ref{sec:results:lc:changing_L}). In our parameters range, changing the loop elongation $a$ in the semielliptical case has a similar effect to changing the loop length $L$ in the semicircular case.

\subsubsection{Group V: (A)symmetric and narrow heating}
\label{annex:sec:groupV_results}

\renewcommand{\arraystretch}{1.3}
\begin{table*}
\caption{Parameters of the group V models.}
\flushleft          
\begin{tabular}{c c c c | c c c | c | c c c c c}   
\multicolumn{12}{c}{Group V} \\
\hline       
\hline       
\multicolumn{4}{c|}{Initial cool loop} & \multicolumn{3}{c|}{Initial hot loop} &
\multicolumn{1}{c|}{} & \multicolumn{5}{c}{Heating} \\
\hline       
  Name & $T_\mathrm{iCL}$ & $n_\mathrm{iCL}$ &  $P_0$  & Name & $T_\mathrm{iHL}$ & $n_\mathrm{iHL}$ & $L$ &  $H_0$ & $A_{\mathrm{max}}$ & $\sigma_\mathrm{ev}$ & $s_\mathrm{ev} - L_\mathrm{mid}$ & iCL $\rightarrow$ fHL
\\
\hline
$m_{19}$ & 4.8 & 9.5 & 0.1 & $m_{38}$ & 5.8 & 9.3 & 3.0 & 0.0023 & 0.79 & 0.3 & 0 & Yes  \\
$m_{20}$ & & & 0.1 & $m_{39}$ & & & & 0.0023 & &  & -1.2 & Yes  \\

\end{tabular}
\tablefoot{Initial parameters of the group V models, with a symmetric and asymmetric localized heating. The parameters are the same as for Table  \ref{table:code:groupI}.   } 
\label{table:code:groupV}      
\end{table*}

\begin{figure*}
    \centering
    \includegraphics{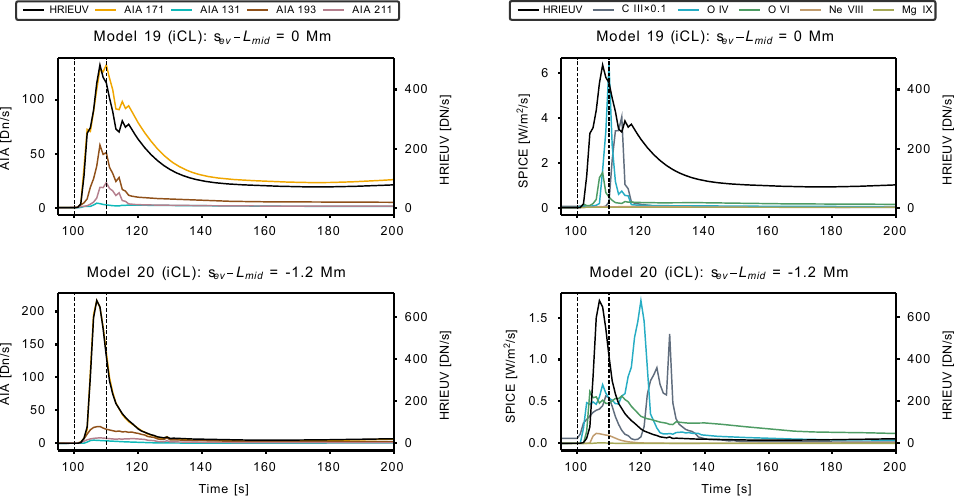}
    \includegraphics{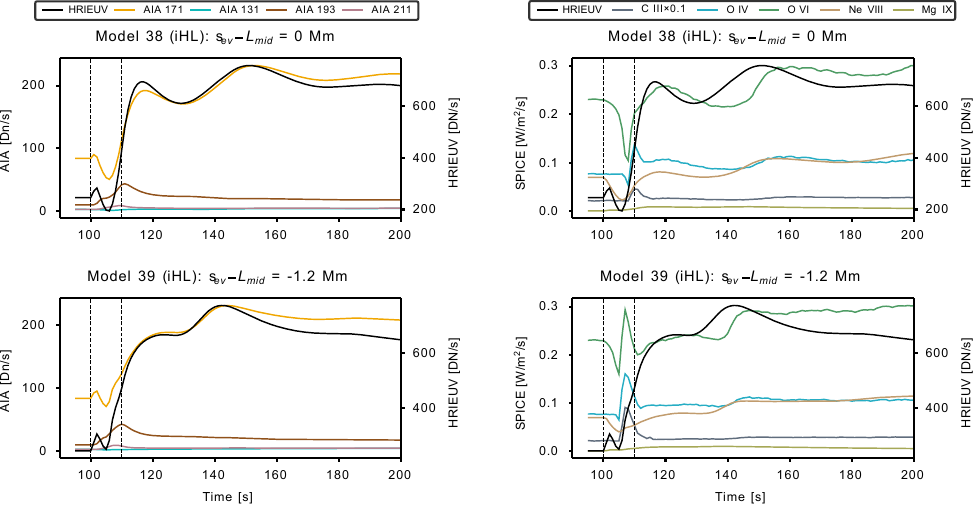}
    \caption{Light curves for the simulations of group V models (Table \ref{table:code:groupV}), with loops in an initial cool or  hot state. The figure is similar to Fig.\ref{fig:results:lc:CL_HL_groupI_these}}
    \label{fig:annex:group5_light_curves}
\end{figure*}

We present the results for the group V models (Table \ref{table:code:groupV}), with an impulsive heating following a narrow deposition function, either located on the loop center (m$_{19}$, m$_{38}$) or on the left side (m$_{20}$, m$_{39}$). 

Figure \ref{fig:annex:group5_light_curves} displays the light curves for all models. Regarding model m$_{19}$, all light curves have an intensity peak with less than \SI{10}{\second} of delay. These results are similar to other models with loops in an initial cool state. m$_{20}$, on the other hand, has delays up to \SI{30}{\second} between the \ion{O}{vi} and the \ion{C}{iii} light curves. When the center of the deposition function is not localized on the loop apex, delays can appear between the intensity peaks of the SPICE light curves. As for models with loops in an initial hot state (m$_{38}$ and m$_{39}$), we notice more than \SI{30}{\second} delays between the HRIEUV and AIA 193 intensity peaks. As such, the results are similar to other models with loops in an initial hot state.

\section{Existence criteria for cool loops}
\label{sec:annex:profile_cl}

In this appendix, we discuss the existence criteria on cool loops. Assuming that the radiative loss function follows a power law $\Lambda(T)\propto T^b$, and that the atmospheric heating follows $H_0 \propto n^{\gamma} $, the condition $\gamma < 2$ implies that cool loops in equilibrium can only exist if $b > 2$ for  $T <$ \SI{e5}{\kelvin} \citep[][]{Antiochos_1986,Cook_1989}. Taking this condition into account, \cite{Cally_1991} argued that the existence of cool loops in the solar atmosphere is unlikely, because their equilibrium condition $b > 2$ cannot be reached. Indeed, the authors referred to the work by done \cite{Athay_1986}, who reviewed the correction of the optical depth effects applied by \cite{McClymont_1983}. They concluded that the exponent $b=3$ for $\Lambda(T)$ is over-estimated, and a more realistic exponent should be closer to $b=2$ for temperatures below \SI{1e5}{\kelvin}. 

However, \cite{Peter_2004} simulated the existence of cool and dense structures emitting at chromospheric to lower temperatures, using a 3D MHD code developed by \citep{Gudiksen_2002,Gudiksen_2005a,Gudiksen_2005b}. The authors used an optically thin radiative loss function without any correction for the optical depth effects. These cool and dense structures were similar to the low-lying loops (below \SI{5}{\mega\meter}) observed in IRIS and AIA, which are episodically heated to TR temperatures ($T \sim$ \SI{e5}{\kelvin}), \citep[][]{Hansteen2014}. This provides evidence that structures similar to the cool loops described by \cite[][]{Antiochos_1986} are likely to exist in the solar atmosphere. Furthermore, \cite{Sasso_2012} have thoroughly evaluated the equilibrium conditions of cool loops over a wide variety of models for the radiative loss function. The authors showed that cool loops in a "quasi-static" state could exist for many models of radiative loss functions, including power laws with exponents equal to $b = 2$ and $b = 3$, and radiative loss functions computed by the atomic database and code CHIANTI \citep[][]{Dere_Chianti}, version 6 \citep[][]{Dere_2009}. The authors defined "quasi-static" cool loops as those with oscillating temperature and density profiles that do not evolve into instability. These results are an argument in favor of the existence of cool loops in a quasi-static or equilibrium state in the solar atmosphere. 

In this work, we aim to understand the observational signatures of low-lying loops submitted to impulsive heating. Thus, we chose to use the radiative loss function derived by \cite{McClymont_1983} to easily create cool loops in equilibrium with a variety of loops and heating models (Section \ref{sec:code:groups_models}).

\section{Forward modeling}

\begin{figure*}
    \centering
    \includegraphics{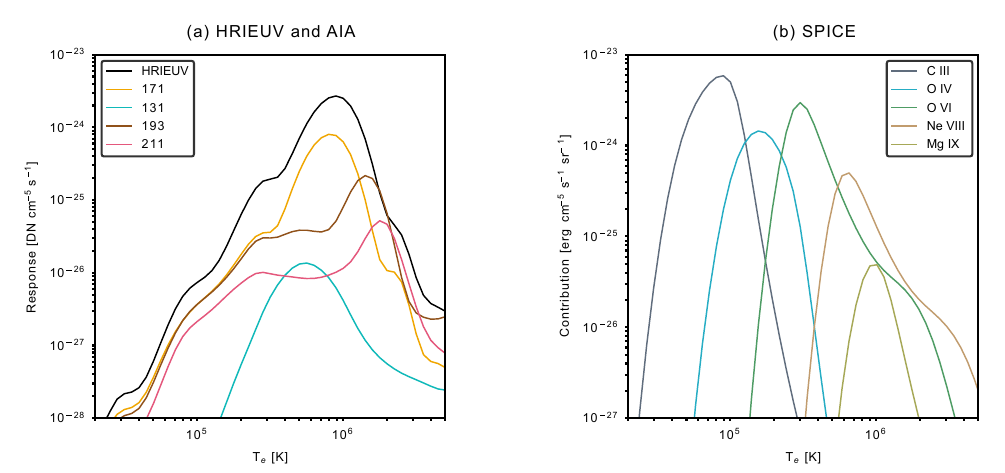}
    \caption{Response functions of (a) HRIEUV, four EUV channels of AIA and (b) and contribution functions of five emission lines measured by SPICE, computed with CHIANTI V 10.1. The abundance is set to ones estimated in the corona by \cite{Asplund_2021} and the ionization equilibrium is the one recommended by CHIANTI. Here, the density is fixed to $n_\mathrm{e}=$ \SI{1e9}{\per\centi\meter\tothe{3}}.}
    \label{fig:annex:gofnt}
\end{figure*}

\label{sec:annex:forward_modelling}
\begin{figure*}
    \centering
    \includegraphics{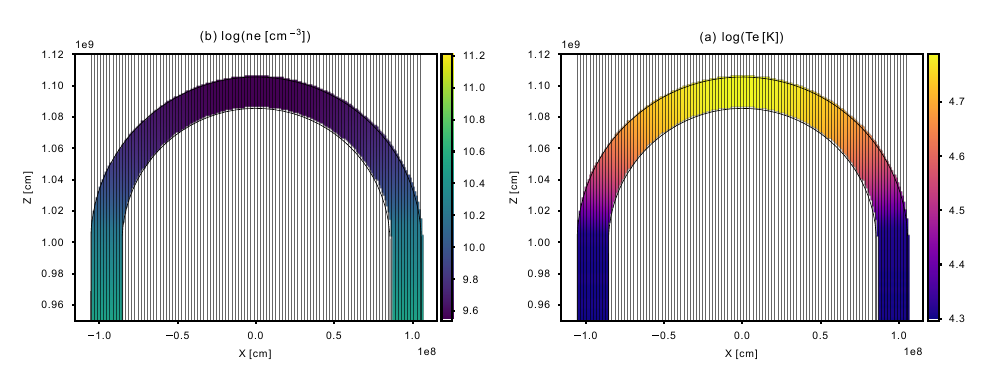}
    \caption{Electron temperature $T_{\mathrm{e}}$ and density $n_{\mathrm{e}}$ from the initial cool loop in equilibrium of the m$_4$ model in the 2D plane ($x$, $z$), described in section \ref{sec:code:loop_geometry}. The cross-section diameter is constant and set to $h=$ \SI{0.2}{\mega\meter}. The vertical black lines are the LOSs chosen to compute the HRIEUV, AIA and SPICE intensities. They are uniformly separated along $x$ by \SI{20}{\kilo\meter}.}
    \label{fig:annex:m4_Tene}
\end{figure*}

In this section, we discuss the method used to compute the light curves for HRIEUV, four EUV channels of AIA and five emission lines measured by SPICE. The intensities are derived from the simulations outputs of electron temperature and density profiles along $s$, along with the response and contribution functions of each channel and lines (Fig. \ref{fig:annex:gofnt}). 

The forward modeling method requires us to build 2D maps of the electron temperature and density at each time step, in order to integrate the intensity along lines of sights (LOS). To do this, we use the ($x, z$) coordinates defined in Section \ref{sec:code:loop_geometry}. The diameter of the loop cross-section is assumed to be constant for all models and equal to $h=$ \SI{0.2}{\mega\meter}.

Figure \ref{fig:annex:m4_Tene} shows the electron temperature and density 2D maps for the initial cool loop in equilibrium of the m$_4$ model, on a ($x$, $z$) grid. The pixel size is equal to \SI{10}{} and \SI{1}{\kilo\meter} along $x$ and $z$, respectively. The LOS are defined as vertical and uniformly separated by \SI{20}{\kilo\meter} along $x$. The intensities $I$ are obtained by integrating the emission along the LOS:

\begin{equation}
\label{eq:annex:I}
    I = \int_\mathrm{LOS} n_\mathrm{e}^2(l) G(T_\mathrm{e}(l), n_\mathrm{e}(l)) \ \mathrm{d}l
\end{equation}

We assumed quasi-neutrality ($n_\mathrm{e} \approx n_\mathrm{i}$). The contribution function $G$ (Fig. \ref{fig:annex:gofnt}b) is used to compute the intensities of the lines measured with SPICE. For imagers (HRIEUV and AIA), $G$ is being replaced by the response function $R$ (Fig. \ref{fig:annex:gofnt}a). In both cases, $G$ and $R$ contain the relevant atomic physics to derive the line emissivity or the channel intensity. They are computed using the atomic database and code CHIANTI \citep{Dere_Chianti}, version 10.1 \citep{Del_zanna_chianti_2021,Dere_2023}. We assume the photospheric abundances estimated by \cite{Asplund_2021}, and the ionization equilibrium recommended by this version of CHIANTI. For the EUV channels of AIA, we take into consideration the instrument degradation estimated on 2022 March 8, as it was the date of the main event dataset from \cite{Dolliou_2024}. The degradation values were given by the \software{aiapy \citep{Barnes2020}\footnote{\url{https://gitlab.com/LMSAL_HUB/aia_hub/aiapy}, consulted on 2024 August 5.}} open project.

At each time step, deriving Eq. \ref{eq:annex:I} results in an intensity for each LOS and each channels or lines. The light curves are directly obtained by averaging the intensity over all of the LOS for each time steps of the simulations. 

\section{Additional figures}

\begin{figure*}
    \centering
    \includegraphics{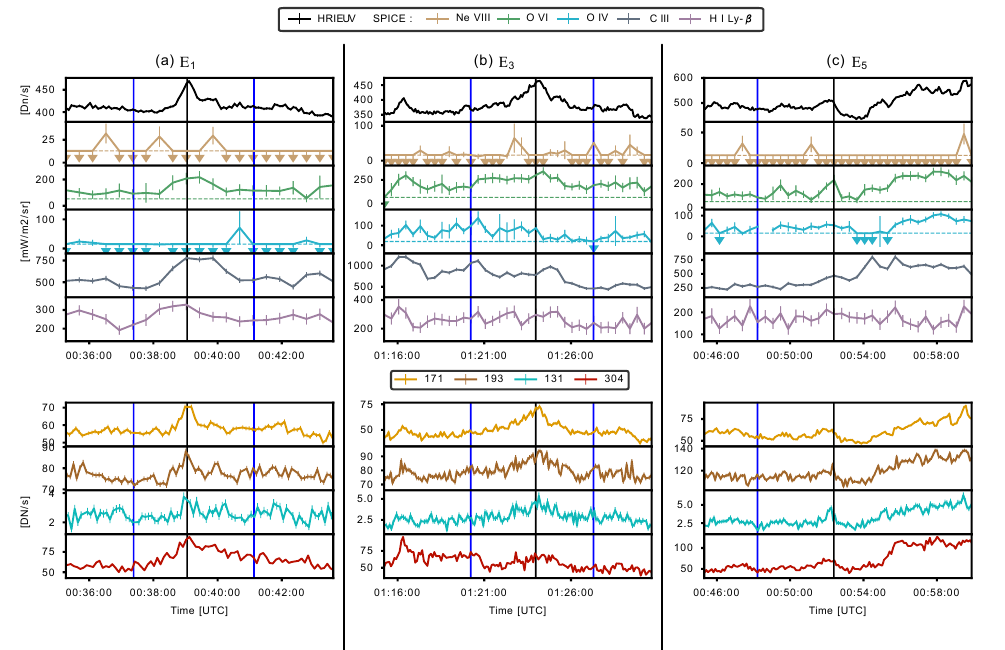}
    \caption{Reproduction of Fig.6 from \cite{Dolliou_2024}, with permission from the authors. The light curves of HRIEUV, the five lines measured by SPICE, and the four AIA channels are given for three EUV brightenings called \evmethod (a), \evd (b), and \evb (c). The vertical black lines indicate the SPICE times closest to the HRIEUV peaks. Detailed information about the vertical black and blue lines are given in the original paper. }
    \label{fig:annex:fig6}
\end{figure*}

\begin{figure*}
    \centering
    \includegraphics{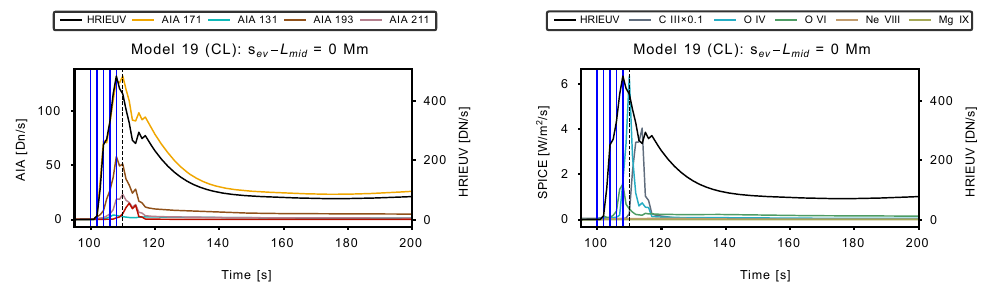}
    \includegraphics{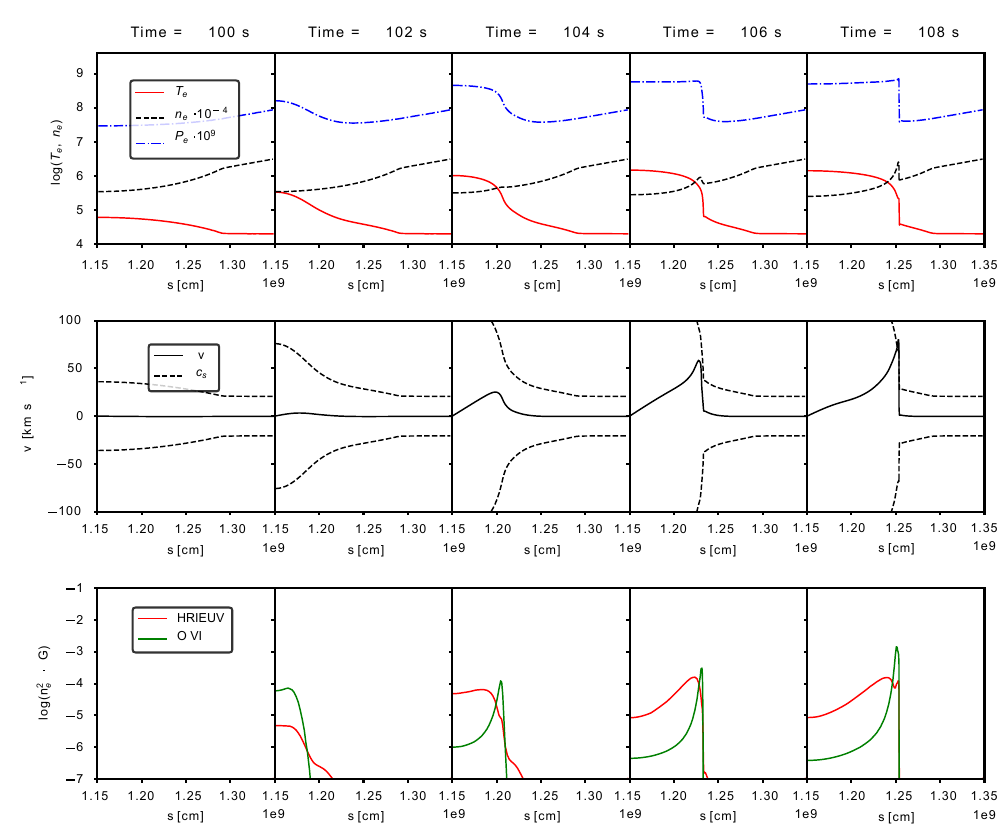}
    \caption{Same as for Fig.\ref{fig:results:CL_HL_m4_lc_ntv}, zoomed-in to the right side of the loop, for the model m$_{19}$ (Table \ref{table:code:groupV}). This model consists of an initial hot loop submitted to an impulsive heating with a narrow deposition function ($\sigma_{\mathrm{ev}} = 0.1L$) localized at the center of the loop ($s_{\mathrm{ev}} = L_\mathrm{c} + L/2$). }

    \label{fig:annex:CL_HL_m19_ntv_v5}
\end{figure*}

\begin{figure*}
    \centering
    \includegraphics{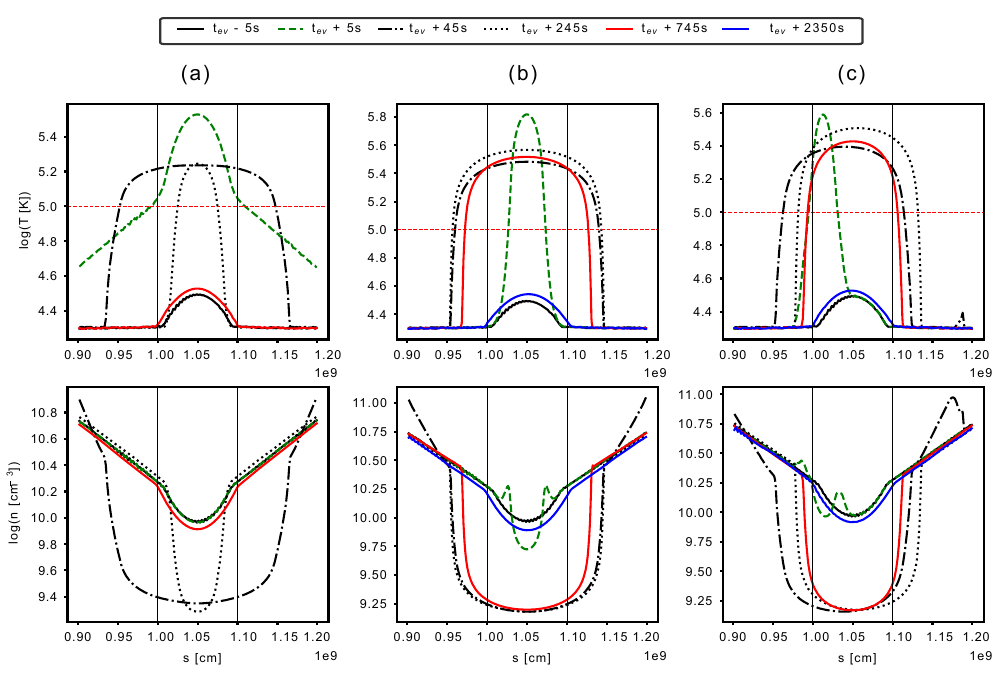}
    \caption{Evolution of the electron temperature (top row) and density (bottom row) profiles for two models with the initial cool state of m$_{2}$ ($L = $ \SI{1}{\mega\meter}). Each column corresponds to an impulsive heating with the following parameters: (a) $A_{\mathrm{max}} = $ \SI{0.5}{\erg\per\centi\meter\tothe{3}\per\second}, $\sigma_{\mathrm{ev}} = 2L$ and $s_\mathrm{ev} = L_{\mathrm{c}} + L/2$ ; (b) $A_{\mathrm{max}} = $ \SI{1.0}{\erg\per\centi\meter\tothe{3}\per\second}, $\sigma_{\mathrm{ev}} = 0.1L$ and $s_\mathrm{ev} = L_{\mathrm{c}} + L/2$ ; (c)  $A_{\mathrm{max}} = $ \SI{1.0}{\erg\per\centi\meter\tothe{3}\per\second}, $\sigma_{\mathrm{ev}} = 0.1L$ and $s_\mathrm{ev} = L_{\mathrm{c}} + 0.1L$. The starting time of the impulsive heating is set at $t_{\mathrm{ev}} = $ \SI{100}{\second}. The black vertical lines indicate the basis of the loop central part and the horizontal dotted red line shows the value $\log{T} = 5.0$. }
    \label{fig:annex:v5_referee_nt_evolution}
\end{figure*}

\end{appendix}
\end{document}